# Learning Motifs and their Hierarchies in Atomic Resolution Microscopy

**Authors:** Jiadong Dan[1,2,3,†], Xiaoxu Zhao[4,†], Shoucong Ning[3], Jiong Lu[5], Kian Ping Loh[5], Qian He[2], N. Duane Loh[3,6,7*], Stephen J. Pennycook[1,2,*,‡]

**Affiliations:**

[1]NUS Graduate School for Integrative Sciences and Engineering, National University of Singapore, 21 Lower Kent Ridge, 119077, Singapore.

[2]Department of Materials Science and Engineering, National University of Singapore, 9 Engineering Drive 1, 117575, Singapore.

[3]NUS Centre for Bioimaging Sciences, National University of Singapore, 14 Science Drive 4, 117557, Singapore.

[4]School of Materials Science and Engineering, Nanyang Technological University, Singapore, 639798, Singapore.

[5]Department of Chemistry, National University of Singapore, 3 Science Drive 3, Singapore, 117543, Singapore.

[6]Department of Physics, National University of Singapore, 2 Science Drive 3, Singapore, 117551, Singapore.

[7]Department of Biological Sciences, National University of Singapore, 16 Science Drive 4, Singapore, 117558, Singapore.

*Corresponding authors: Emails: duaneloh@nus.edu.sg (N.D.L); stephen.pennycook@cantab.net (S.J.P).

[†]These authors contribute to this work equally.

[‡]Present address: School of Physical Sciences, University of Chinese Academy of Sciences, Beijing 100049, China.

**Abstract**

**Characterizing synthesized materials to atomic resolution and first-principles structure-property prediction are two critical pillars for accelerating functional materials discovery. However, we are still lacking a rapid, noise-robust, framework to extract statistically significant multi-level atomic structural motifs from complex synthesized materials to complement, inform, and guide our first-principles models. Here, we present a machine learning framework that rapidly extracts a hierarchy of increasingly complex structural motifs from atomic resolution images. We demonstrate how such motif hierarchies can rapidly reconstruct specimens with vacancies, dopants, domain boundaries, and high-disorder topological features. Abstracting complex specimens with simplified motifs enabled us to discover a new structure in a Mo-V-Te-Nb polyoxometalate (POM), and quantify the relative disorder in a twisted bilayer $MoS_2$. Additionally, these motif hierarchies provide statistically-grounded clues about the favored and frustrated pathways during self-assembly. The motifs and their hierarchies in our framework coarse-grain disorder in a manner that allows us to understand a much broader range of multi-scale samples with functional imperfections and nontrivial topological phases.**

**Teaser**

An interpretable machine learning framework learns structural motifs and their compositional hierarchies in complex materials, which reveal their formative dynamics, intermediate states, and their boundaries of order and disorder.

# MAIN TEXT

## Introduction

Physical systems are often modeled as an ensemble of recurring motifs. These include atomic structural features within periodic unit cells, atomic point defects(*1–4*), extended line defects(*5–9*), topological phases(*10–13*), and superlattices(*14–16*). The composition and spatial arrangement of these motifs underpin the physics of materials. In these cases, rationally-inspired discovery of novel phases and structures using first-principles approaches can be impractical, especially when they contain extended non-periodic features. Such examples are abundant in the rapidly developing field of twistronics, where twisting two monolayers of certain materials can make the resultant moiré bilayer into a superconductor(*17*), Mott insulator(*18*), magnet(*19*), or trap light as solitons(*20*). Another example is POM, which comprises structural motifs of transition metal oxyanions, that can form a diverse range of three-dimensional (3D) enabling frameworks for catalysis, memory storage, and even nanomedical applications(*21, 22*). For such complex systems, one often turns to exploratory high-resolution microscopy to discover novel structures in the laboratory.

Automatically discovering interpretable unseen structural motifs from atomic resolution images remains an unsolved challenge. Consequently, our ability to rapidly describe novel and complex structures from high-resolution but noisy, incomplete images is severely handicapped. This in turn breaks the real-time feedback for efficiently seeking relevant regions in complex samples, and for making timely decisions on sample characterization and preparation.

Machine Learning (ML) models can be trained to recognize specific types of structural motifs(*23–29*) presented at a particular range of resolution, rotations, translations, and noisy imaging conditions. In many cases, training these supervised models on experimental measurements still requires a sufficiently large corpus of labeled data, which often come from laborious labeling by a human expert or idealized forward models (*e.g.*, simulators) with ground truth. However, it is a well-known issue(*30–32*) that applying these ML models to slightly different samples and/or presentations can produce wrong predictions. Furthermore, while feature-extraction has been automated for some atomic resolution micrographs(*33–35*), in some cases these output features are abstract and not readily interpretable as structural motifs(*33, 36*).

Clustering is an established and powerful form of unsupervised learning that does not require labels. These models rapidly yield feature classes that are readily interpretable. Furthermore, such clustering can be noise-robust when feature classes are formed by signal averaging noisy and incomplete observations of potential class members. Yet these unsupervised ML models do not efficiently model the spatial context of the derived features. Just like this paragraph is not merely a "bag-of-words"(*37*), a high-resolution sample is more than a collection of structural motifs. The spatial context surrounding each motif, and the arrangement of neighboring motifs are crucial. Such spatial context gives us crucial information about how these motifs self-assemble, and shed light on mechanisms that encourage or frustrate them.

For such context-aware motif learning, we are inspired by artificial deep neural networks' (DNN) powerful ability to model complex relationships between features: Convolutional Neural Networks can learn spatial relationships between nearby pixels in a hierarchical fashion(*38, 39*); and Natural Language Processing models can learn compositional rules of words, phrases, and fragments(*40, 41*). However, training such supervised DNNs needs labeled data, which is typically derived from laborious manual labeling of atomic resolution micrographs. The label-free alternative using unsupervised DNNs learns features that are not optimized to be human-interpretable, and hence not primed for insightful co-discovery with humans.

Without using DNNs, it is possible to augment a clustering-based unsupervised classifier with the ability to learn hierarchical relationships between structural motifs. Hence, the representations learned by such a model are readily interpretable as motifs and their contextual hierarchies. Here we adopt a *classify-then-compose* framework that analogously decodes the fundamental motifs ("building blocks") of a micrograph using unsupervised learning, from which more complex motifs are hierarchically and interpretably constructed. Such a bottom-up approach can be useful for context-aware learning of unseen, complex structural motifs with minimal to no supervision. The learning objective is twofold: rapidly obtain human-interpretable features for structural motifs obtained from a noisy atomic resolution image, and rapidly characterize the compositional rules of these motifs.

**Results**

The first step of this framework is to extract structural motifs within the sample, some of which might be previously unknown. Importantly, periodic atomic structures can be described by a finite number of atom-centered motifs (Fig. S1). In realistic samples and imaging conditions, however, each of these motifs will have additional internal parameters that describe imaging uncertainties (*e.g.* measurement errors), strain fields, or other latent factors (Fig. S2 and Fig. S3). Complex motifs can in turn be constructed from simpler motifs, to describe longer-range order within complex samples, and systematically create a taxonomy for classifying increasingly complex motifs.

We illustrate the four steps of our learning framework using the annular dark-field scanning transmission electron microscopy (ADF-STEM) image of monolayer MoSe$_2$ in Figure 1A. The first step is *patch extraction*, where we programmatically extract all fixed-size image patches that are centered on all the visible atom columns (Fig. 1B) by seeking regions of high local contrast or symmetry. The second step is *feature extraction*, where the features in these patches are projected onto the Zernike polynomial (ZP) bases (Fig. 1C). In the third step, these Zernike features are grouped into different motifs using a multistage force-relaxed (FR) clustering algorithm (Fig. 1D to 1G), which we describe below (also see Methods and Materials, Movie. S1). And finally, a hierarchy of increasingly complex motifs is automatically constructed (Fig. 1H), from which the original sample is reconstructed (Fig. 1I). This hierarchy building towards more complex motif combinations is elaborated in Figure 2.

The second step of this framework (Fig. 1C) uses ZPs to reduce the representation of the configurations and shapes of the atom columns in pixelated image patches. This is in contrast to the dimensionality reduction for motif building by Belianinov *et al.*(*34*), where only atomic column configurations were preserved. This Zernike representation offers three key advantages. First, the completeness and orthogonality of ZPs guarantee that any square-integrable function on the unit circle can be decomposed into a linear combination of ZPs with coefficients named Zernike moments without redundancy (Fig. S4). In Figure 1C these Zernike moments ($A_p^q$) are uniquely indexed by either the (*p,q*) tuple, or using the equivalent single index *j* (Table S1). Second, ZPs decomposition effectively rejects uninformative high spatial frequency noise (Fig. S5). We also observed that ZPs are demonstrably more efficient than other methods in reconstructing the ground truth (Fig. S6) and improving clustering performance (Fig. S7 and Fig. S8) with different types and amounts of noise. Finally, rotational symmetries within image patches are self-evident in the Zernike representation, which can be selectively turned on or off to determine relationships between class averages (compare Fig. 1D and 1F against Fig. 1E and 1G). Hence, Zernike features are robust against rotational uncertainty. Our time and space complexity analysis (Fig. S9) also shows that computation Zernike moments via matrix approximation (Fig. S10) are about 7.8 times faster and 1.5 times more memory efficient than PCA (scikit-learn implementation).

Classifying noisy unseen motifs together based on similarities between their dominant Zernike features creates average motifs with even less noise, making downstream labeling more robust.

Automatically creating such class averages is critical in the pre-processing of very noisy cryo-electron micrographs(*42*) and x-ray diffraction patterns(*43*). To do this classification flexibly we introduce a force-relaxed (FR) clustering scheme, which generalizes the efficient uniform manifold approximation and projection (UMAP) scheme as a multistage force-based clustering algorithm (Fig. S11, Movie. S1). Following the comparison workflow in Fig. S12, our FR scheme shows comparable performance with t-SNE(*44*) and UMAP(*45*) with balanced datasets (Fig. S13), and outperforms these two with imbalanced datasets (Fig. S14). The resultant FR layout (Fig. 1F and 1G) can be flexibly adapted to datasets with either discrete motif classes or manifolds of motifs by tuning the relative strengths and schedules of these forces (see Materials and Methods).

Our image patches typically enclose the neighboring atomic columns. This is an efficient way of extracting single-atomic column defects without supervision, which readily appear at the FR layouts (Fig. 1F and 1G). These motif labels that are automatically derived from this classification, plus the relative spatial locations between pairs of motifs, can be hierarchically composed to discover simple relationships between structural motifs (Fig. 1H).

A hierarchy of the structural motifs found in Fig. 1 can be automatically constructed with the steps described in Fig. 2. This construction is most readily done when these motifs fall within a lattice. We perform a second round of unsupervised classification where cells on this lattice are classified by their motif compositions (Fig. 2B). The resultant *motif-cells,* which deviate from the perfect crystal, are in turn programmatically connected to similar adjacent motif-cells (Fig. 2B, 2C). A hierarchy can be automatically formed from these connected (but spatially isolated) motif-cells, where the relationships between and occurrence rates of these motif-cells become apparent (Fig. 2D). Details are discussed in the Methods section.

The analyses above only require several seconds on a modest desktop computer (Table S2). Hence, it can be easily adapted to give live feedback at electron microscopes.

This framework readily generalizes to crystalline samples with higher defect densities (Figure 3). An example of such is a monolayer $WS_2$ doped with Fe and Te atoms (Fig. 3A), where the FR-clustering of the Zernike features rapidly identifies more motif classes (Fig. 3D). Using only the spatial relationships between these motifs, more complex motifs can be hierarchically composed (Fig. 3E): Fe-centered defects that are surrounded by three S columns, which are in turn fenced in by W-columns; Te-centered motifs that are surrounded by three W-columns, which are in turn fenced in by S-centered columns. The original ADF-STEM micrograph can be readily reconstructed with these motifs (Fig. 3B and 3C), where motifs of the background crystalline $WS_2$ monolayer are hidden for clarity. A hierarchy of motif-cells, similar to but more complicated than that in Fig. 2, can be reconstructed for this sample (see Fig. S15).

Our framework can create powerful annotations for understanding structures that are too disordered for manual classification. An example of this is the large family of POMs, where multiple polyhedral transition metal oxyanion units link together to form complex three-dimensional structures that hold promise for novel catalytic(*21*) and biomedical applications(*22*). Figure 4A to 4C, illustrates how our framework rapidly identified three types of pentagonal motifs in the Mo-V-Te-Nb-oxide POM and their relative abundances. These motifs resemble five $MO_8$(M=Mo, Nb) octahedra arranged in a pentagon but with the following differences: 59.4% of these pentagons are empty; 36.3% likely surround either niobium (Z=41) or molybdenum (Z=42) atomic columns; and 4.3% possibly surround vanadium (Z=23) atomic columns (Fig. S16).

Remarkably, these three fundamental pentagonal motifs tessellate the plane resembling a type-4 monohedral pentagonal structure(*46*) (Fig. 4D and 4E). This tessellation, which canonically involves only a single species of irregular pentagon, is instead supported here by a spectrum of subtly distorted pentagons (Fig. S17).

Hierarchically composing higher-level motifs from the two major types of single-pentagon motifs (Fig. 4F, Fig. S18) reveals how these pentagonal units might have spontaneously assembled. Level 2 motifs comprise four pentagons that occupy the corners of a larger square; here the putative Vanadium-centered pentagons (light blue motifs in Fig. 4C) are omitted. The blue arrows directed towards the corners of these level 2 motifs indicate if the pentagons at the specified corners are filled. In a completely disordered sample, one would expect $2^4$=16 such level 2 motifs that occur with equal probability. Yet Figure 4F shows that a lower configurational information entropy structure where >83% of these level 2 motifs are dominated by only 6 of these motifs. This low-entropy signature suggests the preferential attachment of pentagonal, but distorted, units during their solution-phase growth(*21*).

Even larger level 3 square motifs can be formed where each corner is now occupied by a level 2 square motif. Again, in a completely random and disordered structure, there should be $2^{16}$=65,536 level 3 motifs of equal probability. And yet again, nearly a quarter of the field of view (Fig. 4G, Fig. S19) is dominated by only two distinct level 3 motifs that conspire to form a previously unreported ordered structure with an approximate 3:1 ratio. From Figure 4G it is clear that the growth of this new ordered structure was entropically frustrated by a large family of competing motifs rich in Mo and Nb columns (Fig. S20). This frustration could have been further abetted by the putatively V-enriched motifs, which appear in the regions between these ordered structures as indicated in turquoise dots (Fig. 4H).

When the variations in the sample are more continuous than combinatorial, constructing motifs can be uninformative. Consequently, building hierarchies from such motifs are potentially misleading. A good example of such samples is the family of van der Waal heterostructures whose electronic and structural properties can be manipulated by introducing relative rotations between atomically thin ordered layers. Figure 5 shows an ADF-STEM micrograph of two hexagonal $MoS_2$ layers that were mutually rotated by approximately 3.15 degrees (Fig. S21), creating a characteristic Moire pattern (Fig. 5A). Although our Zernike-based framework automatically identifies a finite number of high-symmetry structures from the micrograph (Fig. 5B), the continuum of possible relative rotations and shifts between layers causes the experimentally observable variations between these features to be more continuous than discrete in nature. Hence, unlike the FR-layouts in Figures 1, 3 and 4, the FR-clustering map of these Zernike features, which were made rotationally invariant, cannot be meaningfully classified as a finite number of motifs (Fig. 5C).

Despite the continuous variations between the features in this bilayer $MoS_2$ sample, anchor motifs can still be induced amongst them. These anchor motifs, corresponding to the recognizable AA', A'B, AB' idealized domains of this twisted bilayer, are still evident in FR-layout in Figure 5, C and D. Unlike the motifs in Figures 1, 3 and 4, which were created from averages of abundant pseudo-copies of similar features in the sample, the anchor motifs in the twisted bilayer case represent a much smaller fraction of all possible features. Nevertheless, all the features in the micrograph can be quantitatively measured and labeled from these anchor motifs (Fig. 5B). If we separately classify the features closest to these anchor motifs, a hierarchy of motifs emerges describing how AA'-centered, A'B-centered, and AB'-centered domains can be spatially composed (Fig. 5E).

Experimentally realized twisted bilayers will contain imperfections. These imperfections, through the variations between domains or perturbation(*47*), can be readily quantified in our framework. Figure 5, F to G, shows the projected histograms of Zernike features from similar regions centered on two A'B motifs 0 and 1 of Figure 5B. While these two sets of histograms are broadly similar, their visible variations betray differences between these two presumptive A'B domains. Pairwise comparisons of the Kullback-Leibler divergence between these Zernike feature histograms for different A'B domains (Fig. 5H) show divergences (0.184±0.025) that are larger than expected from known nuisance parameters (Fig. S22). Simply put, despite idealized descriptions of these domains(*48*), they contain quantifiably different groups of features.

**Discussion**

Deeply rooted in our framework is the critical notion of structural motifs. Motifs coarse-grain and hence simplify the staggeringly large space of possible atomic configurations in our sample. Each motif class can accommodate variations in features due to uninformative nuisance parameters (*e.g.*, noise, scanning errors, detector noise, *etc.*) or geometric parameters (*e.g.*, small strain fields). With sufficiently high-resolution images, some of these parameters can be unambiguously identified and coarse-grained away. Thereafter, insights about a sample's state and dynamics can be distilled more readily.

This coarse-graining is intimately linked to the fact that interpreting a motif class is more robust against noise and aberrations compared to single noisy image patches. When the signal is sufficiently larger than the noise, such coarse-graining reveals structural motifs (*e.g.*, Fig. 3) despite local strain fields (see POM structure distortion map, Fig. S8). Admittedly, however, for very noisy measurements it will be impossible to discern structural features in the motifs from noise. Additionally, the Zernike polynomial representation offers robustness against the types of scanning drift that were encountered in the experimental images presented here. But again, this robustness will be compromised when this scanning drift becomes sufficiently severe.

These motifs are the essential building blocks for organizing the structural complexity of samples into interpretable hierarchies. In a perfectly periodic structure, the number of atom-centered motifs at any level of this hierarchy equals the number of unique and visible atomic columns within the unit cell (Fig. S1). However, the multiplicity of these motifs will increase with point defects (Fig. S23) and inter-domain boundaries (Fig. S24 and S25). Further introducing random disorder to this periodic structure (*e.g.*, from vacancies, interstitials, translations, and distortions, etc) causes the number of complex motifs to rapidly rise as we build towards more complex hierarchies in our Mo-V-Te-Nb-oxide sample (Fig. S20 and S26). The rate of this rise is a measure of a sample's configurational entropy and loss of long-range order. This entropy also specifies how higher-level motifs can be efficiently and combinatorially represented with lower-level fundamental motifs (Fig. 2, S15, and S27).

The dominant hierarchy of motifs in Figure 3F is strongly influenced by the dominant self-assembly pathways in the sample. In contrast, the sample also reveals a much larger gamut of thermodynamically accessible but less likely higher motifs (Fig. S20 and S26). The exercise of constructing these hierarchies may reveal crucial clues about preferential attachment and free-energy barriers during multi-step nucleation and growth of ordered phases(*49*). Additionally, these motif hierarchies also provide statistically-grounded structural waypoints to guide (or check) *ab-initio* calculations of viable structures, dynamics, and downstream function(*2*, *3*). These ideas can be readily extended to other atomic-resolution structures, such as those collected using scanning tunneling microscopes (Fig. S28).

To conclude, we have described a framework based on Zernike features and force-relaxed clustering to extract and represent motifs from complex atomic-resolution micrographs, and accelerate downstream labeling. Importantly, this framework continues to exploit the spatial context between simple motifs to learn a hierarchical composition of higher motifs that can reconstruct an image from the bottom up. By explicitly inducing this motif hierarchy in a sample, we can quantify and interpret the degree and types of complexity in a sample. Ultimately, these motifs help us coarse-grain disorder and/or complexity in materials to a degree where new knowledge readily emerges and in turn inspires insightful hypotheses. Hence, our techniques offer a novel and rapid approach to extracting multi-level structural information from atomic-level microscopy images and establishing statistically multiscale structure-property links, paving the way to rapid and automatic discovery of next-generation nanomaterials with complex and unknown features.

**Materials and Methods**

Growth of $WS_2$ thin films with Te and Fe dopants

Tellurium (Te) powder was placed into a quartz boat at a temperature of $T(Te) \approx 450\ °C$. $WO_3$ and $FeS_2$ powders were put in a ceramic boat inside the quartz tube at the center of heating zone. A $Si/SiO_2$ substrate with a clean surface was put on the boat. The growth temperature was set at about 800 °C, and the growth time was 30 min. The flow rate of the Argon (Ar) carrier gas was 90 standard cubic centimeters per minute (sccm).

Growth of $MoSe_2$ thin films via MBE

$SiO_2$ substrates were degassed in the same chamber for 1 h and annealed at 500 °C for 10 min. Mo and Se powders were evaporated from an electron-beam evaporator and a Kundsen cell, respectively. During growth, the temperature of the $SiO_2$ substrates were maintained at 500 °C, with a flux ratio between Mo and Se of ~1:10 and chamber pressure kept at ~9 × 10-10 Torr. Monolayer and bilayer MoSe$_2$ can be obtained when the growth temperature is set at 250 °C.

STEM Sample Preparation

As-grown transition metal dichalcogenide (TMDC) films were first identified by optical microscopy. Cu *QUANTIFOIL*® TEM grids were placed onto the target region of TMDC thin films followed by an isopropyl alcohol (IPA)-assisted polymer free lift-off method. The TEM grids were annealed in ultrahigh vacuum chamber (~1×10-9 Torr) at 180 °C for 10 h prior to STEM imaging to eliminate surface contamination.

STEM Characterization

STEM-ADF imaging were carried out on an aberration-corrected JEOL ARM-200F equipped with a cold-field-emission gun at 80 kV if otherwise stated. Two sets of detector acceptance angle were adopted. A higher detector range (68mrad-280mrad) was used for MoSe$_2$ characterization. A lower detector angle range (30mrad-68mrad) was adopted for WS$_2$ imaging for improved contrast of S vacancy sites. A dwell time of 19 µs/pixel was set for scanning imaging mode. HAADF-STEM imaging of the POM structure was performed on UltraSTEM 200 (operated at 200 kV).

Synthetic Data Generation

Synthetic datasets in this paper are used to evaluate popular dimension reduction methods (PCA, t-SNE and UMAP) against the Force-relaxed clustering. We have listed the details of synthetic data in Table S3. Simple synthetic motifs in Fig. S1-2, S4-5 are directly calculated by adding different two-dimensional Gaussian functions. In addition to binary classes of motifs with different *n*-fold symmetry, synthetic bilayer patterns used to evaluate KL divergence (Fig. S13) are generated by convolution with corresponding Gaussian kernels in the Fourier space using sufficient up-sampled grids. Then the generated moiré patterns are then rescaled to adapt to the resolution of experimental data.

Identification of Feature Points

Identification of feature points follows three steps: smoothing, maximum filtering and locating the points. The workflow is detailed in Fig. S29. The key to successfully extracting feature points is to obtain smooth versions of raw images (Step 1). Depending on image quality and image conditions, different smoothing schemes are adopted. For images with high signal-to-noise ratio (SNR), Fourier space filtering was implemented to keep 10% of lowest frequency components. For images with low SNR, Singular Value Decomposition (SVD) based method was applied to the image. We adopted the SVD based method for images in this work unless stated otherwise. The smooth image was dilated by a local maximum filter (Step 2). The feature points are the locations where input image is equivalent to the dilated version (Step 3). In some cases, this identification scheme conservatively overcounts the number of feature points, and features points can be further reduced via symmetry response of Zernike features or selected in Force-relaxed clustering scheme.

Determination of Motif Patch Size

Patch sizes have to be sufficiently large to capture meaningful local symmetries that are persistent in the sample. These symmetries would be efficiently and interpretably encoded in the Zernike representation of these image patches. Empirically, we found this encoding satisfactory when the side-length of the image patches matches the average length of the repeating unit. The latter can be automatically detected using the radial distribution function (see workflow in Fig. S30. Patch sizes smaller than this recommendation tends to offload local symmetry information to the hierarchies that their consequent motifs form, which may not be as readily interpretable.

Dimensional Reduction with Principal Component Analysis

Let the initial set of feature vectors extracted from the raw image be denoted $X = \{x_1, x_2, \cdots, x_i, \cdots, x_n | x_i \subseteq R^m\}$, where $n$ labels the number of features (image patches), and $m$ is length of feature vector that stores Zernike moments. We then use linear principal component analysis (PCA) to identify the main components of covariance between these features. Thereafter, we reduce the features' dimensionality by projecting them into the PCA components that capture the largest feature-feature variations. In this paper, we project all Zernike features ($j \leq 65, m = 66$) obtained from STEM images into the two largest PCA components ($X \mapsto Y = \{y_1, y_2, \cdots, y_i, \cdots, y_n | y_i \subseteq R^{d=2}\}$, where $d$ labels PCA components).

Force-relaxed Clustering

Our two-stage relaxed clustering comprises a *repulsion-dominated* stage, followed by an *attraction-dominated* stage. The repulsion-dominated stage allows adequate separation distance between structural motifs whose features are mutually most dissimilar. The attraction-dominated stage then adjusts the strength of attraction force to make sure each motif cluster is compact, and that clear decision boundaries can be drawn between clusters. The forces used in this paper and others(*44*, *45*, *50*) are described in Table S4 and elaborated in Supplementary Text. The default hyperparameters in the attraction-dominated and repulsion-dominated stages are {α=1, β=1, n=0, m=2} and {α=5, β=1, n=2, m=5} respectively. Although all examples shown in this work use this default set of hyperparameters, users can optimize them for their own imaging needs.

If a feature (*e.g.*, in Fig. 1C) is another feature's k-nearest neighbour (determined once from the feature matrix $X$), attractive forces will pull them together in this PCA-reduced space; otherwise randomly chosen $k'$ non-neighbour pairs are mutually repelled via a stochastic scheme. Exemplary results are shown in Fig. 1F and 1G, where $k = 10, k' = 5$. These forces iteratively move feature-pairs (Y) in the PCA-reduced layout and are inversely proportional to their mutual separation on this plane and weighted by a similarity metric between the feature pairs in their original Zernike space (X). To provide additional control, we alter the relative strength of these attractive and repulsive forces into a repulsion-dominated stage, followed by an attraction-dominated stage. For numerical stability, these forces are gradually relaxed from a maximum in both stages. The hyperparameters involved are illustrated in Fig. S11.

The evolution of the features that resulted in Fig. 1F and 1G reveals how the two-stage relaxed clustering creates decision boundaries between structural motifs. Features suffer a rapid global dilation during the initial repulsion-dominated stage. During this stage, approximate class boundaries quickly emerge while keeping their k-similar features nearby. This dilation is slowed by a force-relaxation schedule. Subsequently, during the attraction-dominated stage similar features coalesce into well-separated centroids without a global contraction of all the features. This ordering of the stages is crucial for creating clear decision boundaries between features, which can be optimized with the hyperparameters in Table. S3, and Supplementary Text. Further comparison is further discussed in Fig. S13-14.

Automatic Construction of Motif-cell Hierarchy

Here we describe the procedure used to programmatically produce the motif-cell hierarchies in Figures 2 and S15. A variant of this approach is used for the POM dataset (Figures S26 and S27).

1. Use the feature points on an image to define cells that tessellate the entire image. This can be done in two ways: automatically locating the lattice vectors in a quasi-periodic sample which gives a regular grid of cells; or a Voronoi construction where the resultant cells might not be periodic.

2. Each of the lowest level structure motifs labeled by FR-clustering are associated with its encompassing cell. This produces *motif-cells* similar to those seen in Fig. 2B.

3. Single-cell motif-cells (e.g., rhombuses in Fig. 2B) are rapidly classified by the set membership of their motifs. These unique single-cell motifs-cells form the lowest-level of the hierarchy that we will build below. In a mostly crystalline sample, the most frequently observed single-cell motif-cells will belong to the periodic unit cells of the crystal. Here we ignore these dominant motif-cells to focus on how non-dominant motif-cells self-assemble.

4. If the cells' positions are described by lattice vectors $i\vec{u} + j\vec{v}$, then each cell's position can be uniquely represented using only its lattice index (i.e., $(i,j)$).

5. Each non-dominant motif-cell is "grown" by automatically connecting it to adjacent motif-cells in their von Neumann neighborhood. A higher-level motif-cell is created when all the neighbors of its constituent motif-cells have all been connected onto this larger motif-cell.

6. These larger, higher-level, connected motif-cells are ordered by levels according to the number of cells each contains (Fig. 2D). Higher-level motif-cells with the same number of cells are compared to remove duplicates up to an overall 90-degree rotation.

7. To build a hierarchy from these motif-cells, we exhaustively check if each lower-level (i.e., smaller) motif-cells is a subset of any of the higher-level motif-cells (accounting for all possible translations, but not rotations). For visual clarity in the hierarchies of Fig. 2D and Fig. S15, edges are only drawn between higher-level motif-cells and their subsetted lower-level motif-cells should the difference in their levels be smaller than some integer threshold.

**Acknowledgments**
We thank J.I. Wong for STEM experimental assistance. **Funding:** S. J. P. acknowledges funding from Singapore Ministry of Education Tier 1 grant R-284-000-172-114, Tier 2 grant R-284-000-175-112 and from the National University of Singapore. N.D.L acknowledges funding support from the National Research Foundation (Competitive Research Programme grant number NRF-CRP16-2015-05), as well as the National University of Singapore Early Career Research Award. Q. He would also like to acknowledge the support by National Research Foundation (NRF) Singapore, under its NRF Fellowship (NRF-NRFF11-2019-0002). X.Z. thanks the support from the Presidential Postdoctoral Fellowship, Nanyang Technological University, Singapore via grant 03INS000973C150. **Author contributions:** J.D, X.Z and Q.H conducted the STEM experiment. J.D developed and designed the framework. S.N advised on the algorithm. J.L, K.P.L and Q.H advised on the experiments. N.D.L and S.J.P supervised the projects. J.D. and N.D.L. wrote the manuscript with input from all authors. **Competing interests:** The authors declare no competing interests. **Data and materials availability:** All data needed to evaluate the conclusions in the paper are present in the paper and/or the Supplementary Materials.

**Supplementary Materials**
Supplementary Text
Figs. S1 to S30
Tables S1 to S4
Movies S1

# Figure 1

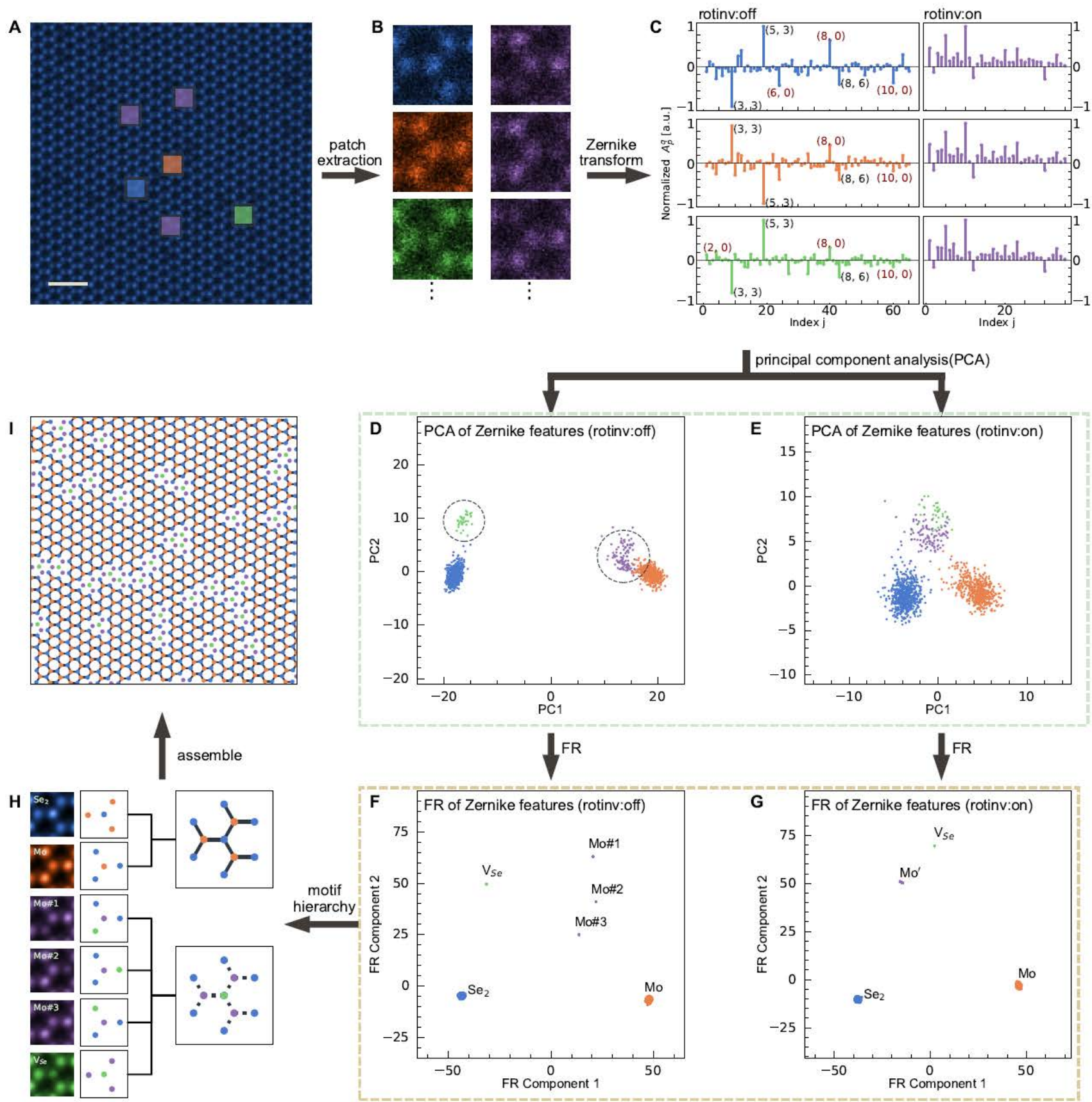

Figure 1. Interpretable machine learning workflow to construct a hierarchy of atomic structural motifs. (A) ADF-STEM image of monolayer $MoSe_2$ (scale bar: 0.5 nm). (B) Automatically extracting atom-centered image patches from panel (A). (C) Computing the Zernike features within the image patches in panel (B) either in their rotationally invariant representation (rotinv:on) or retaining rotation information (rotinv:off). (D) and (E) project the patches' Zernike representations in their first two principal components. (F) Applying a force-relaxed clustering algorithm acting on these Zernike features automatically classifies them into different atomic structural motifs; (G) clustering on the rotationally invariant representations (rotinv:on) neatly groups features centered on Mo atoms together. (H) Each cluster in (F) represents a different structural motif: patches centered on two Se ($Se_2$), single Se vacancies ($V_{Se}$), or single Mo columns that are next to vacancies (Mo#1, Mo#2, Mo#3), and those that are not (Mo). These motifs can in turn be composed as two classes of atom-column quartets. (I) Reconstruction of (A) with these atom-column quartets.

# Figure 2

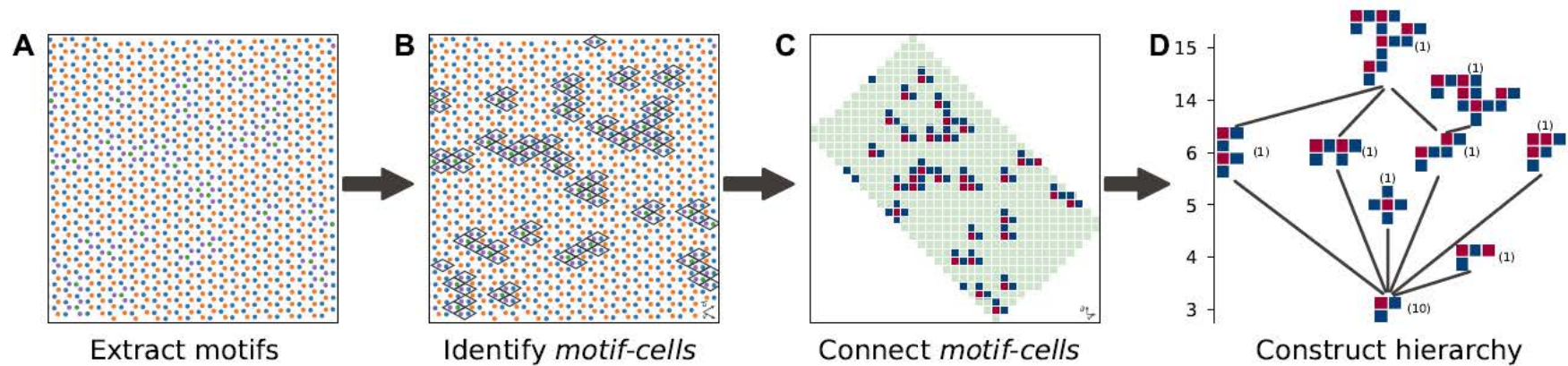

**Figure 2. Steps in the automatic construction of hierarchy of motif-cells found in the monolayer MoSe$_2$ sample.**
**(A)** Lowest level motifs (represented by color dots) from Fig. 1I, and their corresponding positions. **(B)** Bin motifs into motif-cells that are defined by realspace lattice vectors. All motif-cells are classified according to their motif compositions. The minority motif-cells, which correspond to two types of defects, are outlined here as rhombuses. **(C)** Mapping realspace positions of two types of minority motif-cells (blue and purple) into a square grid: spatial locations of cells in (B) shown on a square grid delineated by their cells' realspace lattice indices (rather than vectors). Connected minority motif-cells (von Neumann neighbors) are programmatically identified as higher-level motif-cells. **(D)** These motif-cells are ordered according to the number of cells they contain (enumerated on vertical axis), where differences up to a 4-fold rotation symmetry are ignored. The number of occurrences of each type of motif-cell are shown in parentheses. We associate higher-level motif-cells with lower-level ones if the spatial arrangement of cells in the latter occur within the former; here edges are drawn between associated motif-cells that are the nearest in the hierarchy.

Figure 3

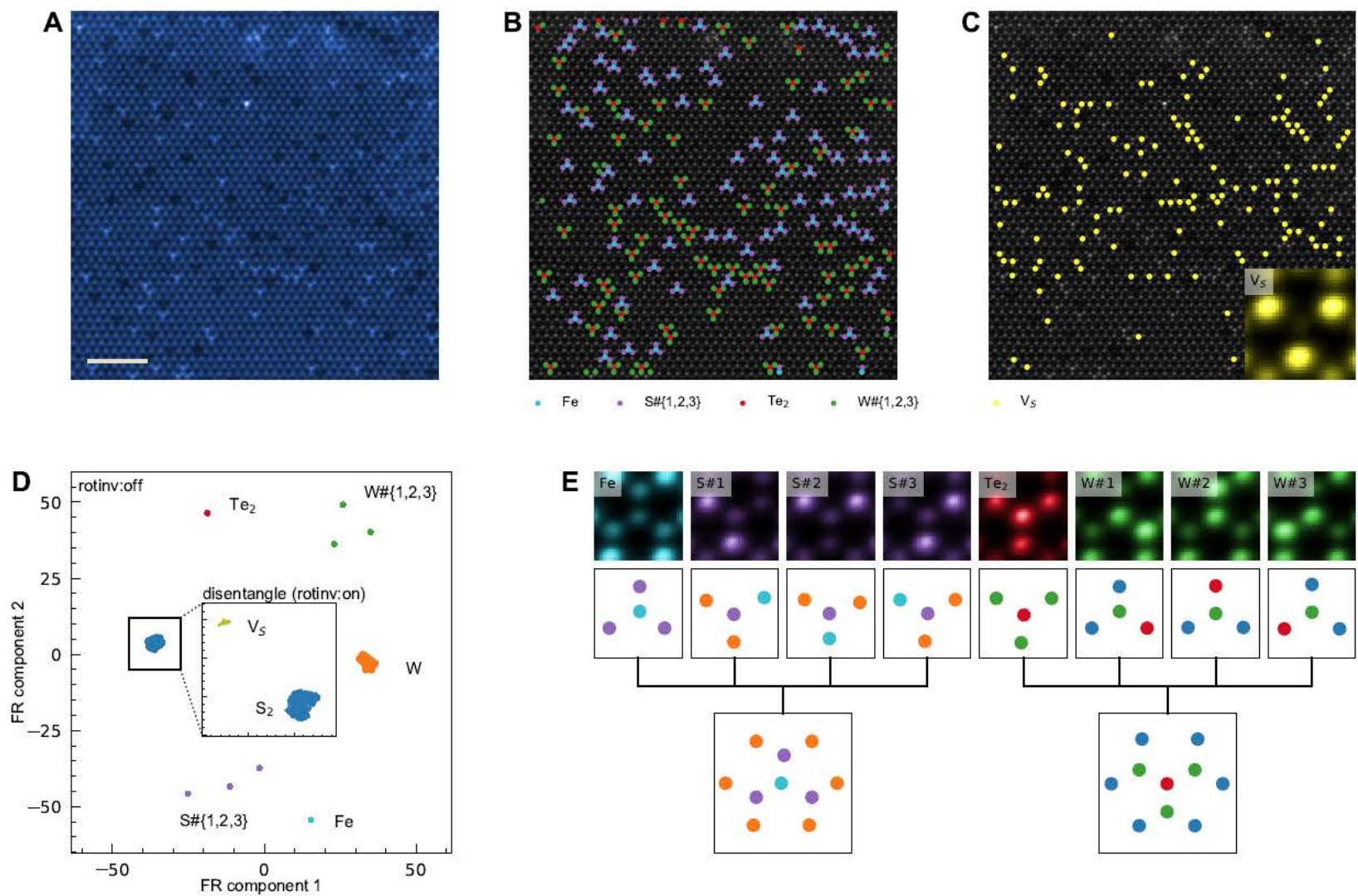

Figure 3. Comprehensive identification of dopant sites in quaternary alloy based on monolayer $WS_2$. (A) Low angle ADF STEM image of monolayer $WS_2$ doped with Fe and Te. Our framework readily identified (B) Fe-centered (turquoise) and Te-centered (red) motifs, as well as (C) single S vacancies $V_S$ (yellow) within the $WS_2$ lattice (gray). (D) FR-clustering of Zernike features (rotation invariance turned off) shows the W-centered motifs (orange) and S-centered motifs (blue), plus the structural motifs that contain dopants (Fe, Te) or have peripheral S or W atom columns (*i.e.*, S#{1,2,3} and W#{1,2,3}). Amongst the motifs comprising only W+S, those that are centered on S vacancies can be identified by turning on rotational invariance (inset). (E) For the motifs containing dopants, their spatial context can be summarized as a tree-like hierarchy with other motifs.

# Figure 4

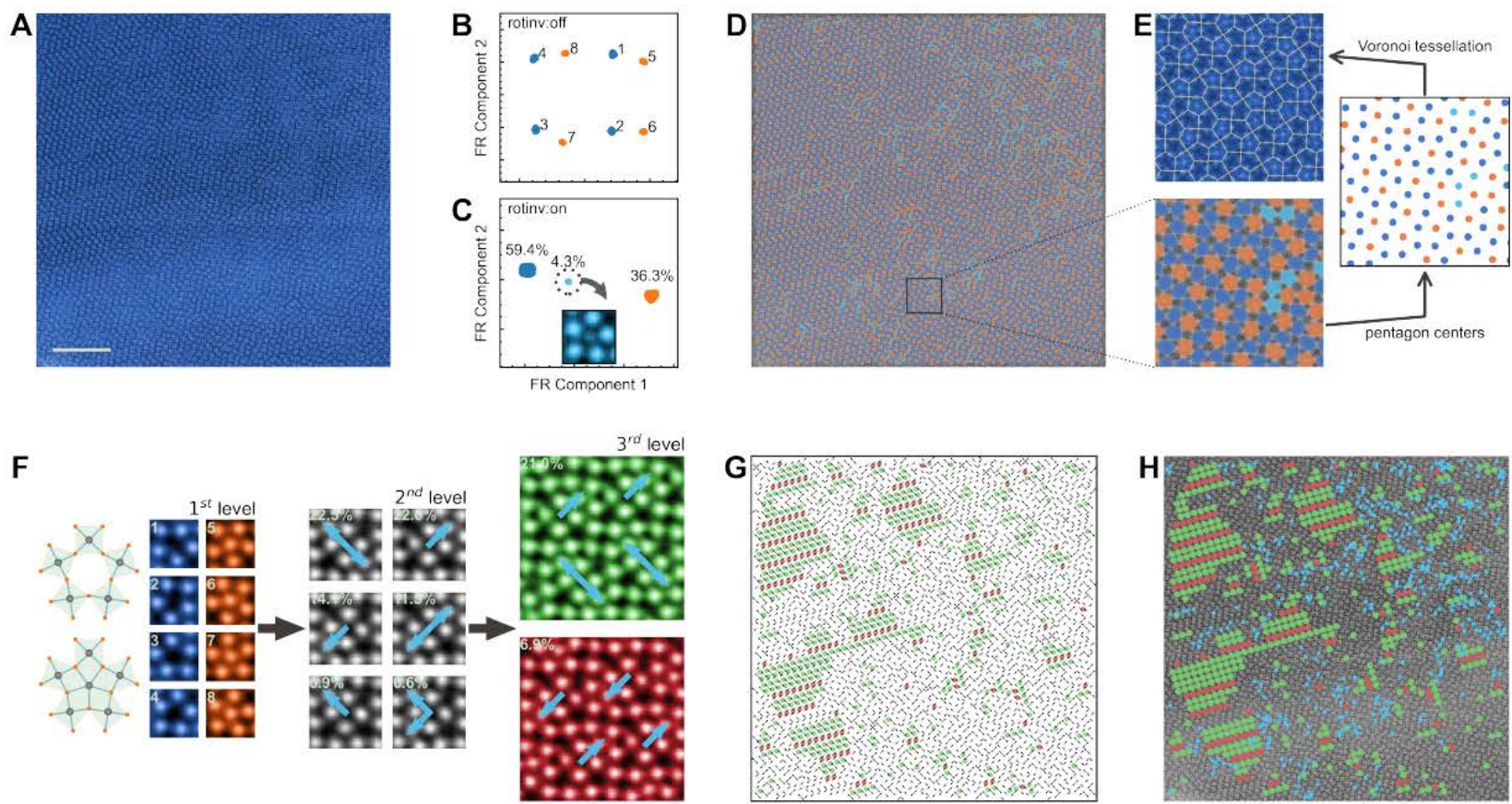

Figure 4. Hierarchy of motifs in polyoxometalate Mo-V-Te-Nb-oxide shows novel but frustrated phase. (**A**) ADF STEM image of the complex metal oxide with no apparent long-range order. (**B**) and (**C**) FR-clusterings of their Zernike features show motifs that are centered on empty atom columns (dark blue), partially filled columns (light blue), or filled pentagonal atom columns (orange). (**D**) While the ADF-STEM image can be reconstructed with these three motifs in a quasi-random way, (**E**) the Voronoi tessellation of their centers form a monohedral pentagonal tiling (type 4). (**F**) Hundreds of larger motifs can be composed from these filled/hollow pentagonal motifs in a 3-level hierarchy (only significant nodes in hierarchy shown here). The most abundant motifs at the third level (green and red) are centered on four-column squares, labeled by arrows indicating the direction of filled pentagons from the centers. (**G**) Reconstruction of the ADF-STEM image with these arrows and (**H**) largest dominant motifs clearly show a novel phase that could tessellate the plane, but is frustrated by other competing structural motifs.

Figure 5

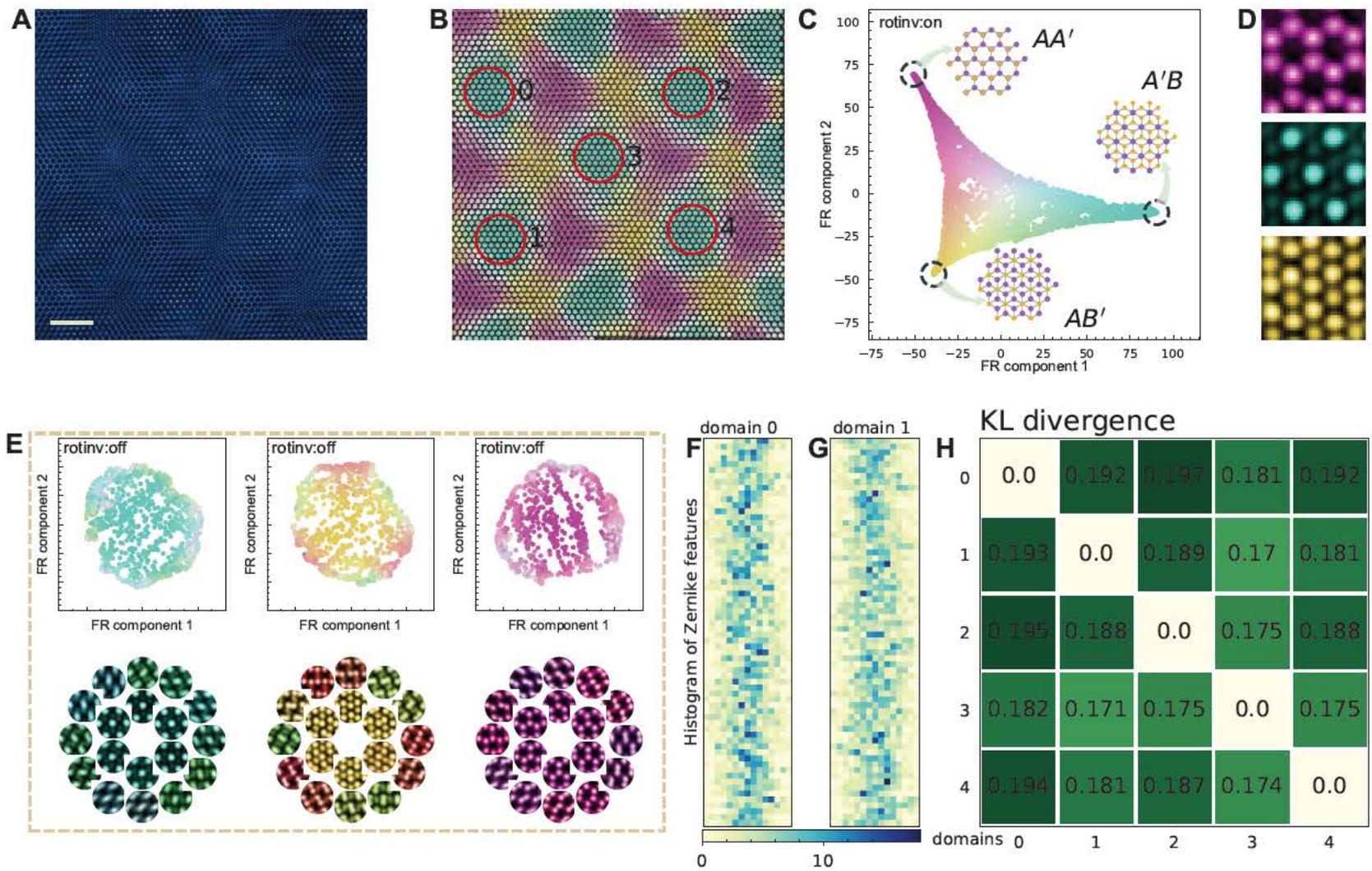

**Figure 5. Reconstructing variations between domains in twisted bilayer MoS$_2$.** (**A**) Low angle ADF STEM image of bilayer MoS$_2$ with 3.15° interlayer rotation (scale bar: 2nm) (**B**) Zernike features in this image, colorized by our framework, show the coexistence of AA', AB', and A'B type domains. (**C**) FR layout of these Zernike features, with the rotational invariant mode turned on. This layout is anchored at its three vertices by the corresponding atomic models from the AA', AB', A'B phases, whose average experimental motifs are shown in (**D**). All the features in (C) are colored based on their distances from these three anchor motifs. (**E**) Top row shows the FR layout of Zernike features from A'B, AB', and AA' domains (left to right) with the rotational invariant mode off. Image patches averaged from the corresponding features in the bottom row show how these features spatially fill the A'B, AB', and AA' domains. (**F**) Binned histograms of all Zernike features in A'B domain 0 in (B) sequentially projected onto its 66 feature dimensions (each shown as a separate row); (**G**) identically binned histogram from features in A'B domain 1 of (B). (**H**) Pairwise Kullback–Leibler divergences between the Zernike feature histograms from the five A'B domains circled in (B).

# Supplementary Materials for

## Learning Motifs and their Hierarchies in Atomic Resolution Microscopy

Jiadong Dan, Xiaoxu Zhao, Shoucong Ning, Jiong Lu, Kian Ping Loh, Qian He, N. Duane Loh, Stephen J. Pennycook

Correspondence to: duaneloh@nus.edu.sg (N.D.L); stephen.pennycook@cantab.net (S.J.P).

**This PDF file includes:**

Supplementary Text
Figs. S1 to S30
Tables S1 to S4
References

**Other Supplementary Materials for this manuscript include the following:**

Movies S1

**Supplementary Text**

Finite Number of Motifs in Crystal sample

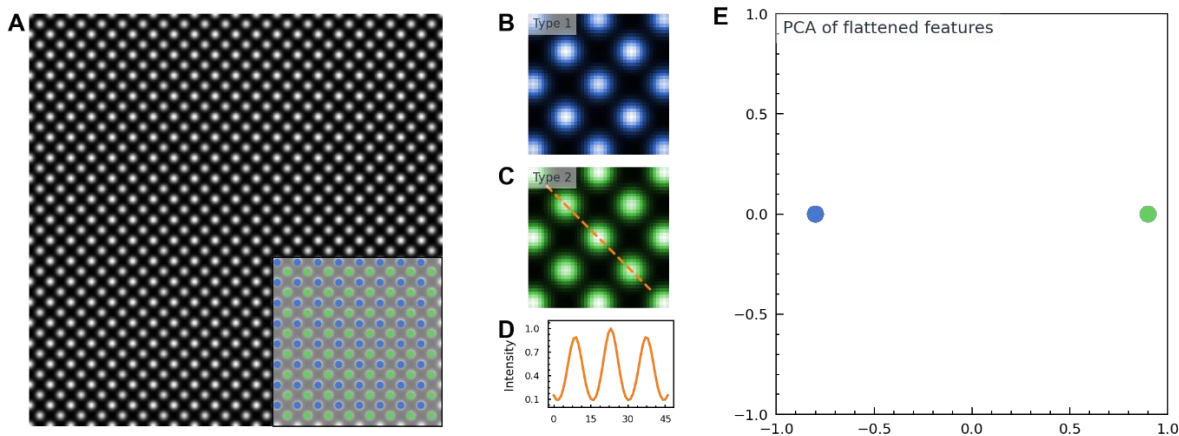

**Fig. S1 Finite number of motifs in a synthetic crystal sample**. (A) Synthetic image of a crystal sample with plane group $p4m$, which comprises two types of motifs: a motif centered with a dimmer blob and a motif centered with a brighter blob as shown in (B) and (C) respectively. (D) Line profile showing the intensity difference in (C). (E) The PCA projection of flattened features demonstrates the discrete nature of this sample.

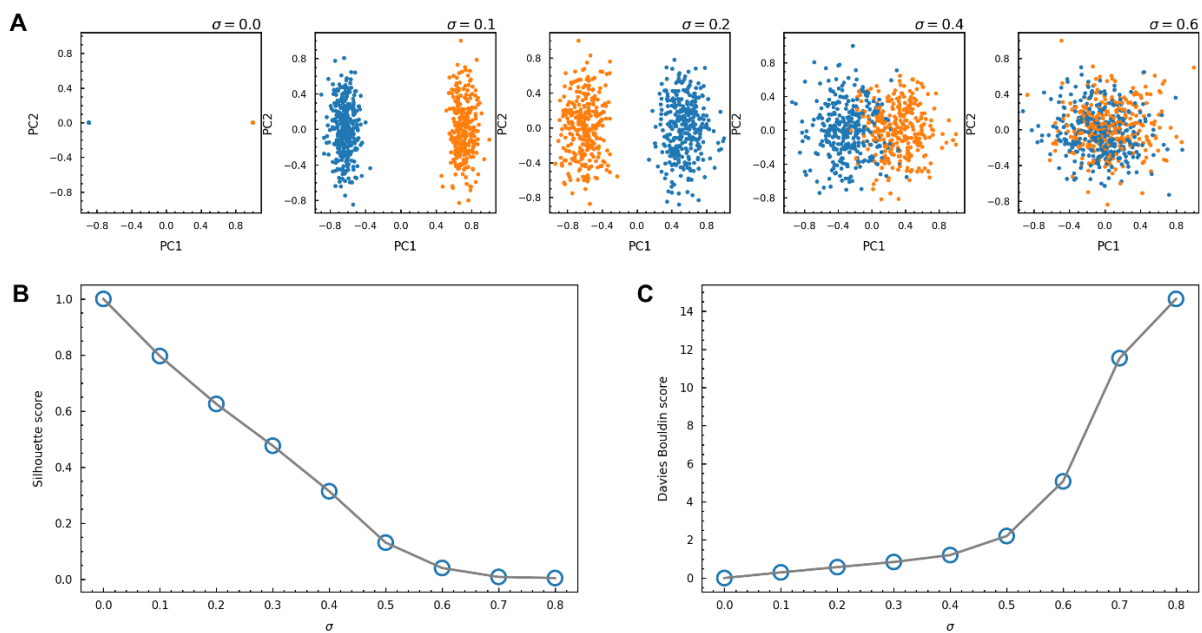

**Fig. S2 Cluster broadening due to addition of Gaussian noise with standard deviation $\sigma$**. (A) The PCA projections of flattened features from the synthetic crystal in Fig. S2 with various Gaussian noise levels. With an increase of $\sigma$ values, two clusters representing two types of motifs

will finally overlap. To evaluate the degree of overlap, the Silhouette score (B) and Davies-Bouldin score (C) are plotted against different $\sigma$ values.

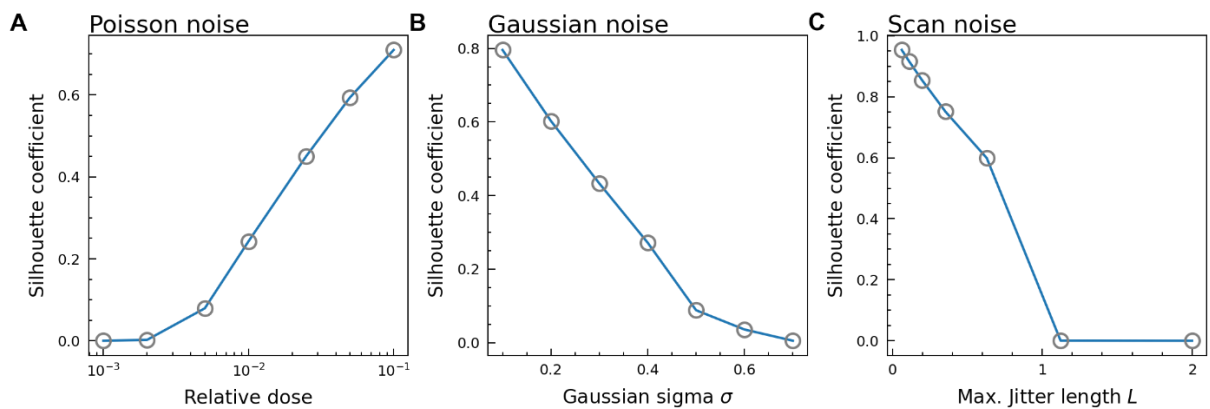

**Fig. S3. Cluster broadening in three different noise settings: Poisson noises, Gaussian noises and scan noises.** Synthetic dataset parameters: $\{n1 = 5000, n2 = 5000, s = 128, \sigma = 7, l = 32, A = 0.8\}$.

Conveniently, STEM images of crystalline samples contain only a finite number of atom-centered motifs, which often correspond to the discrete nature of the low dimensional representation of these motifs. This discrete property is closely related to the fact that crystal structure can be concisely described by a unit cell that only contains a finite number of atoms. We used synthetic data composed of an array of Gaussian blobs to validate this statement. Fig. S1A displays a synthetic image of a crystal sample with plane group $p4m$, which contains two types of motifs as indicated in the inset image and Fig. S1 (B and C). Fig. S1 D shows the intensity difference of two types of blobs. The maximum intensity of the central blob in type 1 and 2 motifs is 0.8 and 1. Fig. S1 E is the PCA projection of flattened features demonstrating the discrete nature of this sample.

Fig. S2 illustrates the broadening of clusters from synthetic data in Fig. S1A with increasing Gaussian noise ($0 \leq \sigma \leq 0.8$). Fig. 2A selectively shows the PCA layouts in different $\sigma$ values. Fig. B and Fig. C depict the Silhouette score and David-Bouldin score against $\sigma$. The decreasing trend in the Silhouette scores, and the increasing trend in the David-Bouldin scores quantitatively demonstrate that these two types of atom-centered motifs are increasingly difficult to differentiate with increasing noise.

Fig. S3 shows the cluster broadening effect is also evident in Poisson noises and scan noises settings. Here we used a synthetic binary class dataset from Table. S3.

Zernike Polynomials and Zernike moments

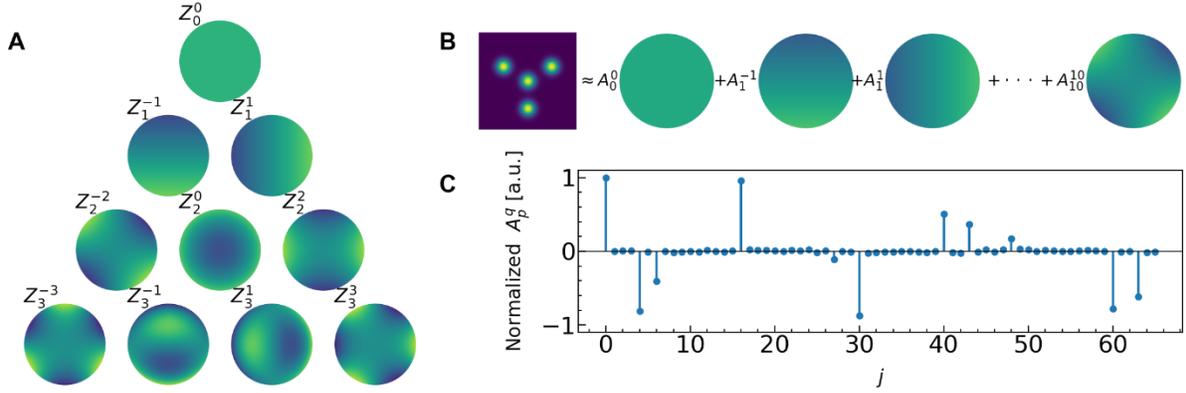

**Fig. S4 Zernike polynomials and Zernike moments**. (A) The first 10 Zernike polynomials arranged in a pyramid form. Each polynomial term is labelled by $Z_p^q$, where $p$ is the radial index and $q$ is the azimuthal index. All Zernike polynomials are vertically arranged by $p$ and horizontally ordered by $q$. (B) The decomposition of image patch to a linear combination of Zernike polynomials. (C) The coefficients for each polynomial can be grouped to form a compact representation of the original patch.

The ZPs are a complete set of orthogonal basis functions defined in the unit disk denoted by the double indexing scheme $Z_p^q(\rho, \theta)$ (Fig. S4), where $p$ is a nonnegative integer, and $q = \{-p, -p+2, -p+4, \ldots, p\}$ for a given $p$. The double indices $(p, q)$ are ordered into a single index $j = (p(p+2) + q)/2$. Each ZP consists of a normalization term $N_p^q$, a radial term $R_p^{|q|}$, and an azimuthal term $sin(q\theta)$ or $cos(q\theta)$:

$$Z_p^q(\rho, \theta) = \begin{cases} N_p^q R_p^{|q|}(\rho) \cos(q\theta); & for\ q \geq 0 \\ -N_p^q R_p^{|q|}(\rho) \sin(q\theta); & for\ q < 0 \end{cases}$$

Here, $R_p^{|q|}$ and $N_p^q$ are given by

$$R_p^{|q|}(\rho) = \sum_{k=0}^{\frac{p-|q|}{2}} \frac{-1^k (p-k)!}{k!\left(\frac{q+|q|}{2}-k\right)!\left(\frac{p-|q|}{2}-k\right)!} \rho^{p-2k},$$

and

$$N_p^q = \sqrt{\frac{2(p+1)}{1+\delta_{q0}}},$$

where $\delta_{q0}$ is the Kronecker delta.

Fig. S4 shows a visual depiction of the first 10 Zernike polynomials. The order of the polynomial is determined by the indices $p$ and $q$. In fact, the double indices $(p, q)$ are ordered into a single index $j$, i.e., $Z_p^q$ is equivalent to $Z_j$. Table S1 shows the conversion between $(p, q)$ to $j$, for which the mathematical relation is expressed below,

$$j = \frac{p(p+2) + q}{2}.$$

The azimuthal component of $Z_p^q$ is either $sin(q\theta)$ or $cos(q\theta)$. In fact, it can be generalized to a complex form using Euler's formula $e^{ix} = cos(x) + isin(x)$. Hence, Zernike polynomials in complex-valued form can be expressed as

$$Z_p^q(\rho, \theta) = N_p^q R_p^q(\rho) e^{iq\theta}.$$

It is noteworthy that $Z_p^q(\rho, \theta)$ is complex valued if derived in complex form, which is useful to prove the rotation-invariance properties of Zernike moments in section <u>Rotational Invariant form of Zernike Moments</u>.

Any square-integrable functions $f(\rho, \theta)$ within a unit disk can be decomposed into an infinite series comprising weighted Zernike polynomials:

$$f(\rho, \theta) = \sum_{p=0}^{\infty} \sum_{q=-p}^{p} A_p^q Z_p^q(\rho, \theta), \quad p - |q| = even$$

where the coefficients $A_p^q$ is can be calculated as

$$A_p^q = \int_0^{2\pi} \int_0^1 f(\rho, \theta) Z_p^q(\rho, \theta) \rho d\rho d\theta.$$

Rotational Invariant form of Zernike Moments

The image function defined in the unit disk $D = \{(\rho, \theta): 1 \leq \rho \leq 1, 0 \leq \theta \leq 2\pi\}$ is denoted by $f(\rho, \theta)$. If a rotation through an angle $\alpha$ is operated on the image function $f(\rho, \theta)$, a rotated version of image function $f^R(\rho, \theta)$ is obtained, and it relates to the original image function by,

$$f^R(\rho, \theta) = f(\rho, \theta - \alpha)$$

The complex-valued Zernike moments of the original image function $f(\rho, \theta)$ is

$$A_p^q = \int_0^{2\pi} \int_0^0 f(\rho, \theta) Z_p^{q*}(\rho, \theta) \rho d\rho d\theta = \int_0^{2\pi} \int_0^0 f(\rho, \theta) R_p^q(\rho, \theta) e^{-jq\theta} \rho d\rho d\theta$$

The Zernike moments of the rotated image function $f^R(\rho, \theta)$ in the same polar coordinate is

$$A_{R_p}^q = \int_0^{2\pi} \int_0^0 f(\rho, \theta - \alpha) R_p^q(\rho, \theta) e^{-jq\theta} \rho d\rho d\theta$$

Let $\theta_1 = \theta - \alpha$,

$$A_{R_p}^q = \int_0^{2\pi} \int_0^0 f(\rho, \theta_1) R_p^q(\rho, \theta) e^{-jq(\theta_1 + \alpha)} \rho d\rho d\theta$$

$$= e^{-jq\alpha} \int_0^{2\pi} \int_0^0 f(\rho, \theta_1) R_p^q(\rho, \theta) e^{-jq(\theta_1)} \rho d\rho d\theta = A_p^q e^{-jq\alpha}$$

This shows that $A_{R_p}^q$ and $A_p^q$ differ by a phase shift that is closely related to the rotational angle $\alpha$. The phase difference of $A_{R_p}^q$ and $A_p^q$ indicates that the magnitudes of complex-valued Zernike moments remain identical to those prior to the rotation, i.e., $\left| A_{R_p}^q \right| = \left| A_p^q \right|$.

Noise rejection of Zernike representations

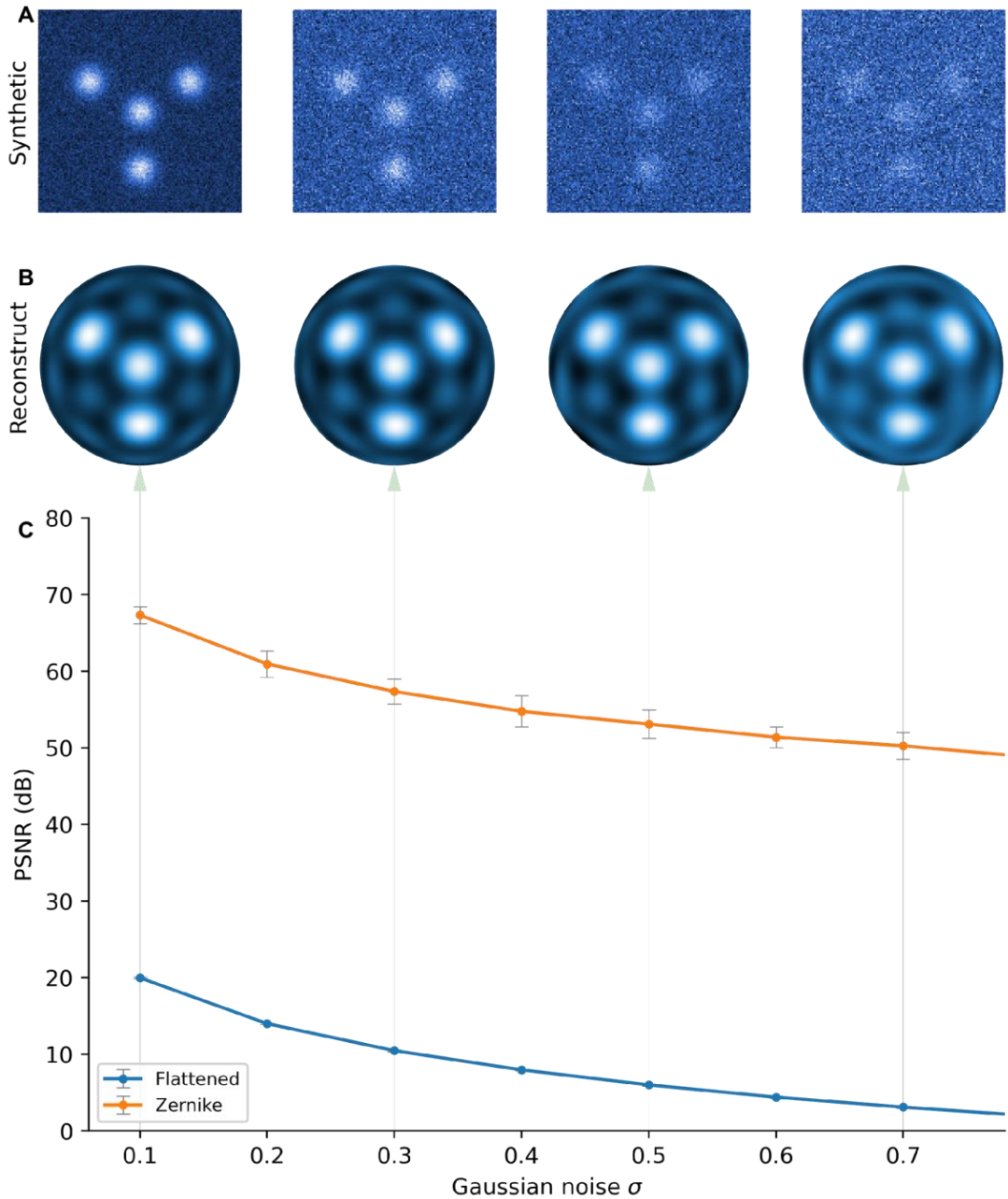

**Fig. S5 Evaluate noise rejection using synthetic patches with 3-fold rotational symmetry.** (A) Synthetic patches with increasing levels of Gaussian noise ($\sigma = 0.1, 0.3, 0.5, 0.7$). (B) The reconstructed patches from the first 66 terms ZPs indicating the main three-fold rotational symmetry even for high levels of noise. (C) Comparison of PSNR of two different feature representations: flattened image representation (blue) and Zernike representation (orange). The Zernike representation shows higher PSNR values in all levels of $\sigma$ values.

The truncated ZP representations can effectively reduce noise. Much of the high spatial frequency measurement noise in high-resolution micrographs is predominantly contained in higher-order Zernike moments. Furthermore, the $p \leq 10$ Zernike projection of an image patch already captures a wide range of possible shapes and arrangements of atomic columns and defects. Therefore, by truncating ZPs beyond *p*=10 effectively allows us to reject higher spatial frequency measurement noise. As shown in Figure S5, with a very low value of PSNR (1.81), the reconstruction patch still shows the main three-fold symmetry which is consistent with the Zernike moments plot. Fig. S5 shows that the low spatial frequency features of noisy images are at least three orders of magnitude more detectable than the putative input peak signal to noise ratio in different symmetry configurations.

Comparison of ZPs and PCA from patch reconstruction

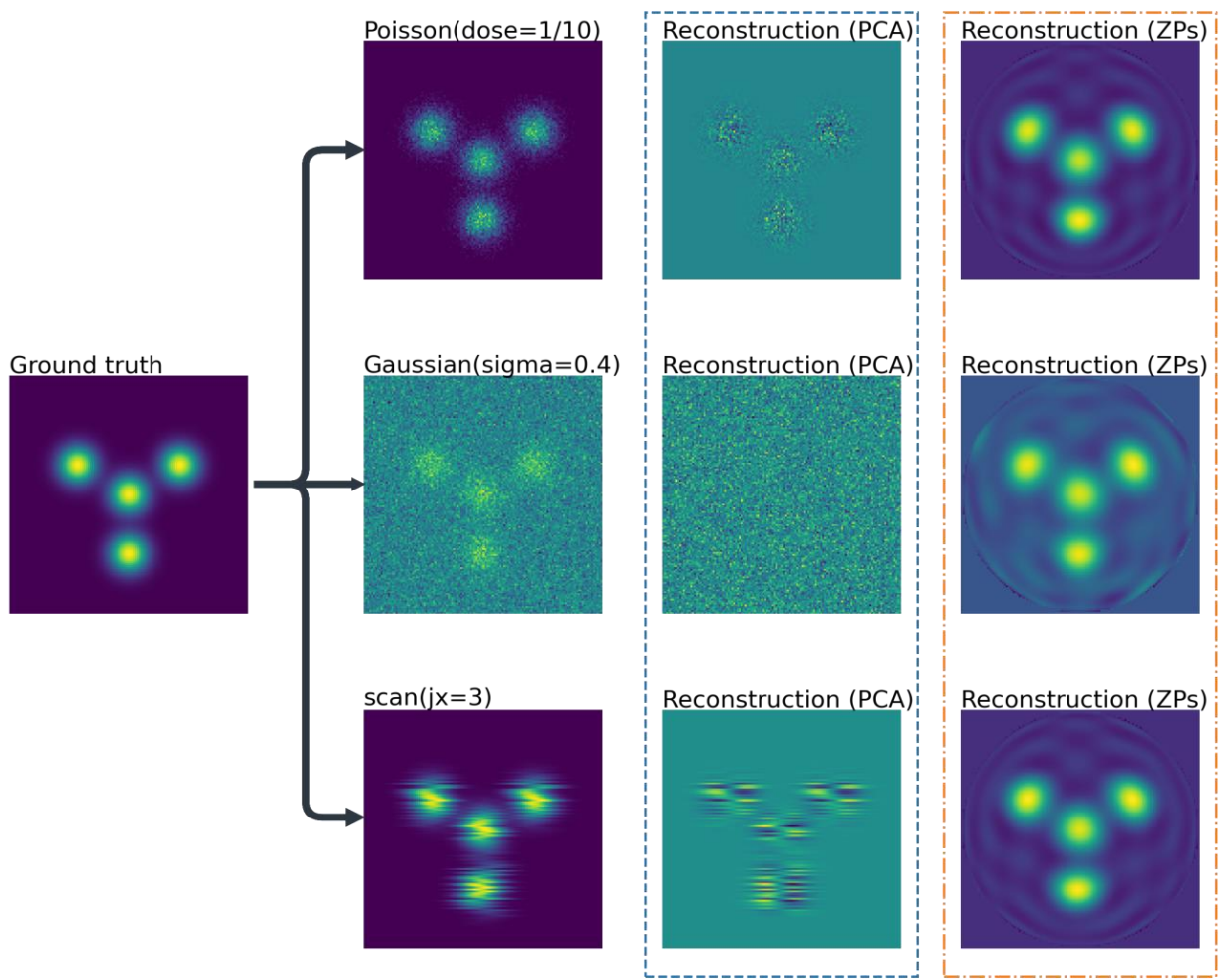

**Fig. S6. Comparison of Reconstruction results using PCA and ZPs in different noise settings.** ZPs reconstructions use the first 136 (corresponding to p=15 and q=15) terms of Zernike

polynomials. PCA reconstructions are computed from 500 stacks of synthetic patches, and truncated using the same number of components as ZPs.

To enrich the discussion towards why we choose ZPs and provide further evidence of the usefulness of our method, we have added more comparison analysis to validate our point. We have basically two paths to compare ZPs and PCA: compare the reconstruction results from PCA and ZPs both truncated to the same number of components; or compare clustering performance of features reduced using PCA or ZPs.

Fig. S6 shows the reconstruction results using PCA and ZPs when truncated to the same number of components in three different noise models. After applying only a moderate noise in all three scenarios, we observed that ZPs (orange rectangle) outperform PCA (blue rectangle) in recovering the spatial location and contrast of the ground truth patches. We have to point out that PCA is an effective image denoising method when implemented in overlapping patch-based singular value decompositions (SVD) method. Poisson noise is significantly reduced when overlapping patches are grouped as input into the SVD-related denoising algorithm(*1*). In representing features from patches using PCA, it contains variations from noise components if pre-pre-processing/denoising is lacking.

Comparison of ZPs and other dimension reduction methods via cluster performances

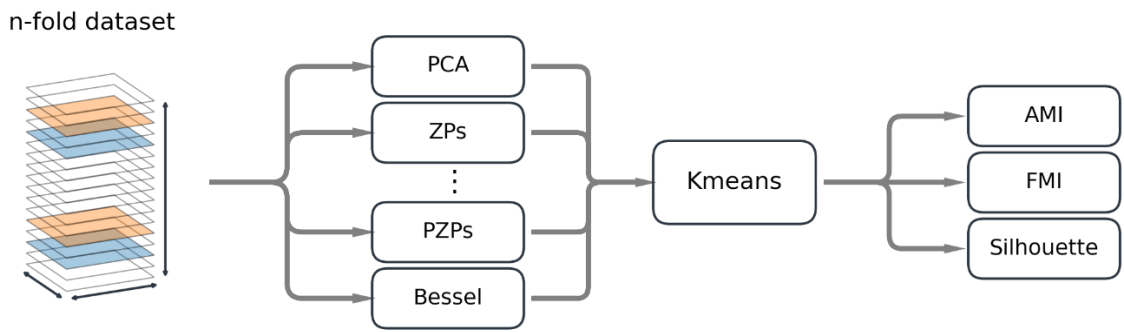

**Fig. S7. Workflow to compare different representation (dimensional reduction) methods via clustering performance scores.**

The advantages of ZPs over PCA are even clearer if evaluated based on clustering performance scores via a scheme shown in Fig. S7. The synthetic dataset (of different noise levels) is firstly represented via different dimension reduction techniques, then the reduced features are clustered using the KMeans algorithm. The predicted labels from Kmeans combined with ground truth labels are used to calculate different clustering performance scores including adjusted mutual information (AMI), Fowlkes–Mallows index (FMI) and Silhouette coefficient.

Fig. S8 shows a complete comparison of these scores in different noise settings using various representation methods. In all three noise models, fixed-bases methods (*e.g.*, ZPs, PZPs(*2*) and Bessel(*3*)) outperforms PCA and kernel PCA. In particular, ZPs show the best clustering performance in Poisson setting.

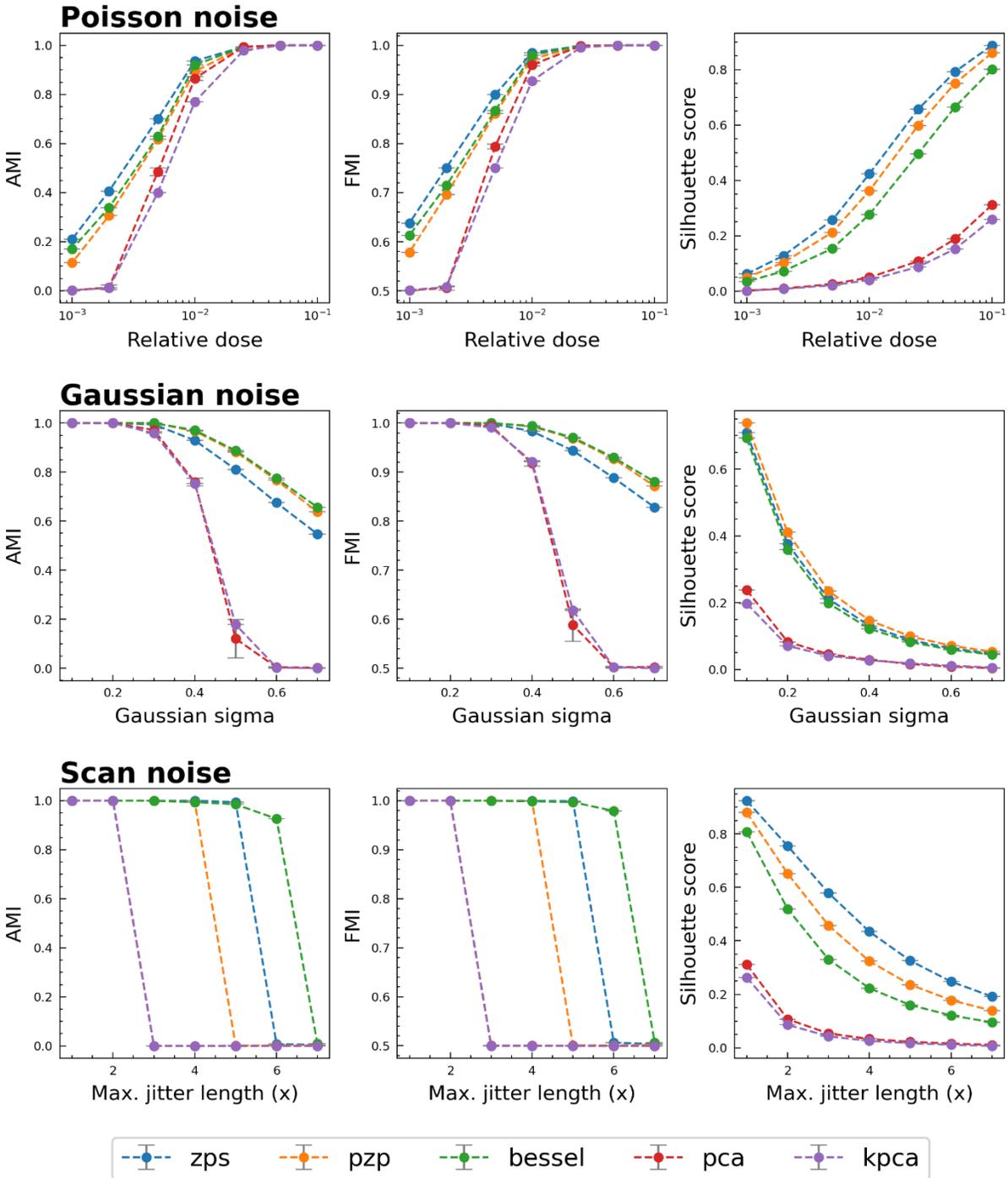

**Fig. S8. Comparison of different representation (dimension reduction) methods.** Top panels: AMI, FMI and Silhouette scores of different representation methods in *Poisson noise* setting. Middle panels: AMI, FMI and Silhouette scores of different representation methods in *Gaussian noise* setting. Bottom panels: AMI, FMI and Silhouette scores of different representation methods in *scan noise* setting.

Time and Memory Scaling Analysis of ZPs and PCA

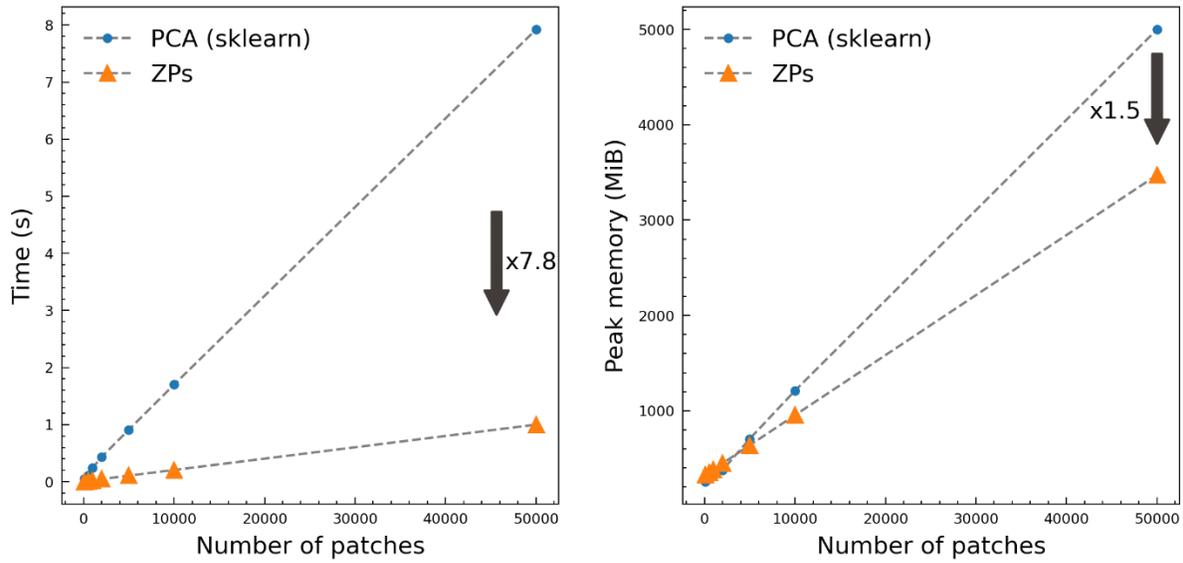

**Fig. S9. Time and space time complexity analysis of PCA and ZPs.**

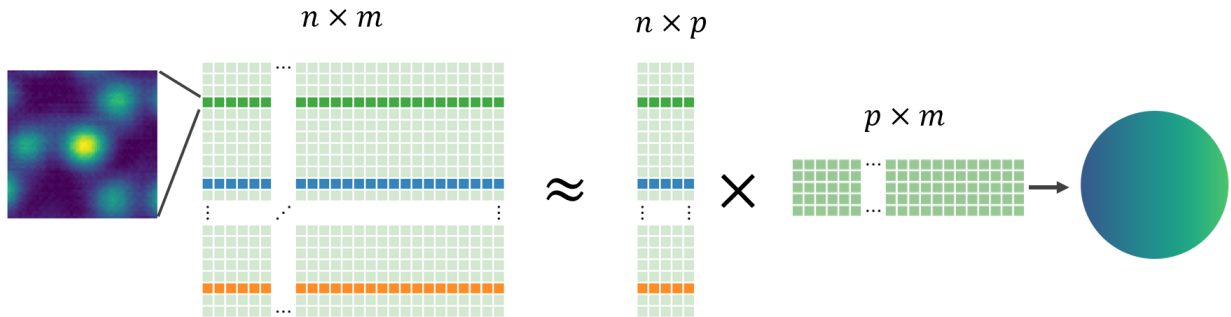

**Fig. S10. Computation of Zernike moments (features) via matrix approximation.** We used matrix approximation to compute Zernike features with a reduced dimension of $p$. A total number of $n$ image patches was flattened to form a matrix with a shape of $(n, m)$, where $m$ is the number of pixels in one patch. A set of $p$ Zernike polynomials with the same shape of image patch is also flattened to form a matrix with a shape of $(p, m)$. The reduced features with a shape of $(n, p)$ can be approximated via matrix pseudo-inverse operation.

Computation Zernike moments can be greatly sped up via matrix approximation instead of directly integrating. As illustrated in Fig. S10, the linear decomposition property of ZPs entitles us to simplify the computation through a matrix pseudo inverse operation. Fig. S9 shows that computation of Zernike moments via matrix approximation are about 7.8 times faster and 1.5 times more memory efficient than PCA (scikit-learn implementation).

Force-relaxed Clustering

1. Basic Definitions

Let $X = \{x_1, x_2, x_3, \cdots, x_n\}$ be a set of $n$ features such that $x_n \subseteq R^m$. In this work, $x_n$ represents the $m$ Zernike moments extracted from the $n^{th}$ image patch. The set of such features $X$ then suffer a second low dimensional transformation $T: X \mapsto Y = \{y_1, y_2, y_3, \cdots, y_n | y_n \subseteq R^d \}$, and $d \ll m$. Here, we use principal component analysis (PCA) with $d = 2$ to generate the initial layout $X \rightarrow Y$. Different low dimensional transformation can also be adopted.

The neighborhood information of $X$ can be stored in a weighted graph $G_X = (V_X, E_X)$, where $V_X$ are vertices (features) and $E_X$ are edges (between valid pairs of features) of the graph. Edges are drawn between two vertices only if they have a non-zero element in their adjacency matrix, $P \in R^{n \times n}$ that we define below. The element $P_{ij}$ ($1 \leq i \leq n$ and $1 \leq j \leq n$) of the adjacency matrix P is the weight of the edge $(i, j)$, which measures the similarity between vertices associated with features $x_i$ and $x_j$.

An isomorphic graph can be constructed with the reduced features $Y$ as vertices: $G_Y = (V_Y, E_X)$. Notice that both graphs $G_X$ and $G_Y$ share the same set of edges.

2. Construction of an Adjacency Matrix P from X Using k-Nearest Neighbor Method

Given a set of vertices $V_X$, there are many ways to construct the edges of graph $G_X(V_X, E_X)$. In this work, we use the $k$-nearest neighbor method ($k$-NN).

Given an input hyperparameter $k$, we are able to compute the set of $k$-nearest neighbors of $x_i$ as $\{x_i^1, \ldots, x_i^k\}$. For each $x_i$, we define a minimum distance $r_i$ and a normalization distance $\sigma_i$ from the neighborhood. The minimum distance

$$r_i = min\{d(x_i, x_i^\kappa) | 1 \leq \kappa \leq k\},$$

where $d(x_i, x_i^\kappa)$ is a default distance metric between two features $x_i$ and $x_i^\kappa$. In this work, $d$ between two features vectors $u$ and $v$ is defined as

$$d(u, v) = 1 - \frac{(u - u^-) \cdot (v - v^-)}{\| (u - u^-) \| \| (v - v^-) \|},$$

where $u^-$ and $v^-$ are the average of the components of feature vectors $u$ and $v$ respectively.

We then choose to normalize the distance metric around different features to homogenize the density of features in $R^m$. This choice, empirically, will cause the features to cluster at approximately similar rates. Specifically, we normalize the distance around each feature $x_i$ with a $\sigma_i$ defined in the equation:

$$\sum_{\kappa=1}^{k} exp\left(-\frac{d(x_i, x_i^\kappa) - r_i}{\sigma_i}\right) = log_2(k).$$

Here $\sigma_i$ is numerically determined using a binary search for the k-neighbors around each feature vector $x_i$.

Then we can construct an asymmetric adjacency matrix $Q$ between all pairs of features.

$$Q_{ij} = \{exp\left(-\frac{d(x_i, x_j) - r_i}{\sigma_i}\right) \text{ if } x_j \in \{x_i^1, ..., x_i^k\} \text{ } 0 \text{ otherwise},$$

where $1 \leq i \leq n$ and $1 \leq j \leq n$.

In this notation, we can rewrite $Q_{ij}$ as,

$$Q_{ij} = \{exp\left(-\frac{d(x_i, x_j) - r_i}{\sigma_i}\right) (i,j) \in E_X \text{ } 0 \text{ otherwise}.$$

The full adjacency matrix $P$ is obtained by symmetrizing $Q$:

$$P = Q + Q^\top.$$

Edges are defined $(i,j) \in E_X$ for the graph $G_X$ only if $P_{ij} \neq 0$. The set of all possible edges complementary to $E_X$ is denoted by $E^-{}_X$. This complementary set will be used later in the repulsion stage of our clustering algorithm.

3. <u>Updating Y According to the Neighborhood Information Stored in P</u>

To update the features in their reduced space $Y = \{y_1, y_2, y_3, \cdots, y_n\}$, we apply attractive forces between vertices $(i,j) \in E_X$ and repulsive forces between vertices in the complementary $(i,j) \in E^-{}_X$. This update is done iteratively, labeled by the iteration index $t$, such that $1 \leq t \leq t_{max}$. Further these force-directed updates in $Y$ are separated into two stages, marked by iteration number $\tilde{t}$: an attraction-dominated stage ($1 \leq t \leq \tilde{t}$), followed by a repulsion-dominated stage ($\tilde{t} < t \leq t_{max}$). This iteration partition $\tilde{t}$ can be tuned by hand; here, we set $\tilde{t} = t_{max}/2$.

Additionally, we also introduce a function $\gamma(t)$ that linearly relaxes the forces from unity to zero during the attraction-dominated and repulsion-dominated stages separately (see Figure S6).

Let us denote the forces between any pair of vertices, $y_i$ and $y_j$, as $f_a(||y_i - y_j||)$ for attractive forces, and $f_r(||y_i - y_j||)$ for repulsive forces. We can update the reduced features from iteration $(t)$, $Y^{(t)} = \{y_1^{(t)}, y_2^{(t)}, y_3^{(t)}, \cdots, y_n^{(t)}\}$, to $(t+1)$ using the following recipe.

1) Given a particular ordering of the edge lists $E_X$, we sequentially attract the pairs of vertices $(i,k) \in E_X$ (e.g. $(1,2), (3,4), (2,1), ...$). Explicitly, we move each attractive pair $(i,k) \in E_X$ symmetrically,

$$y_i^{(t+1)} = y_i^{(t)} - \gamma(t)P_{ik}f_a(\| y_i - y_k \|)\left(y_i^{(t)} - y_k^{(t)}\right) y_k^{(t+1)}$$
$$= y_k^{(t)} - \gamma(t)P_{ik}f_a(\| y_i - y_k \|)\left(y_k^{(t)} - y_i^{(t)}\right).$$

2) After each attraction between two vertices $(i,k)$, we then repel the $i$th vertex against 10 randomly selected non-neighbor $j$ vertices (i.e. $(i,j) \in E^-_X$). This repulsion is also done symmetrically on $(i,j) \in E^-_X$,

$$y_i^{(t+1)} = y_i^{(t)} + \gamma(t)P_{ij}f_r(\| y_i - y_j \|)\left(y_i^{(t)} - y_j^{(t)}\right) y_j^{(t+1)}$$
$$= y_j^{(t)} + \gamma(t)P_{ij}f_r(\| y_i - y_j \|)\left(y_j^{(t)} - y_i^{(t)}\right).$$

During the attraction-dominated stage ($1 \leq t \leq \tilde{t}$), the customized attractive and repulsive forces $f_a(.)$ and $f_r(.)$ are respectively:

$$f_a = \frac{\alpha}{1 + \| y_i - y_j \|^N} \quad f_r = \frac{\beta}{1 + \| y_i - y_j \|^M}.$$

Then when we switch over to the repulsion-dominated stage ($\tilde{t} < t \leq t_{max}$), the parameters in these forces are replaced with their tilde version (i.e. $\alpha \to \tilde{\alpha}, \beta \to \tilde{\beta}, N \to \tilde{N}, M \to \tilde{M}$).

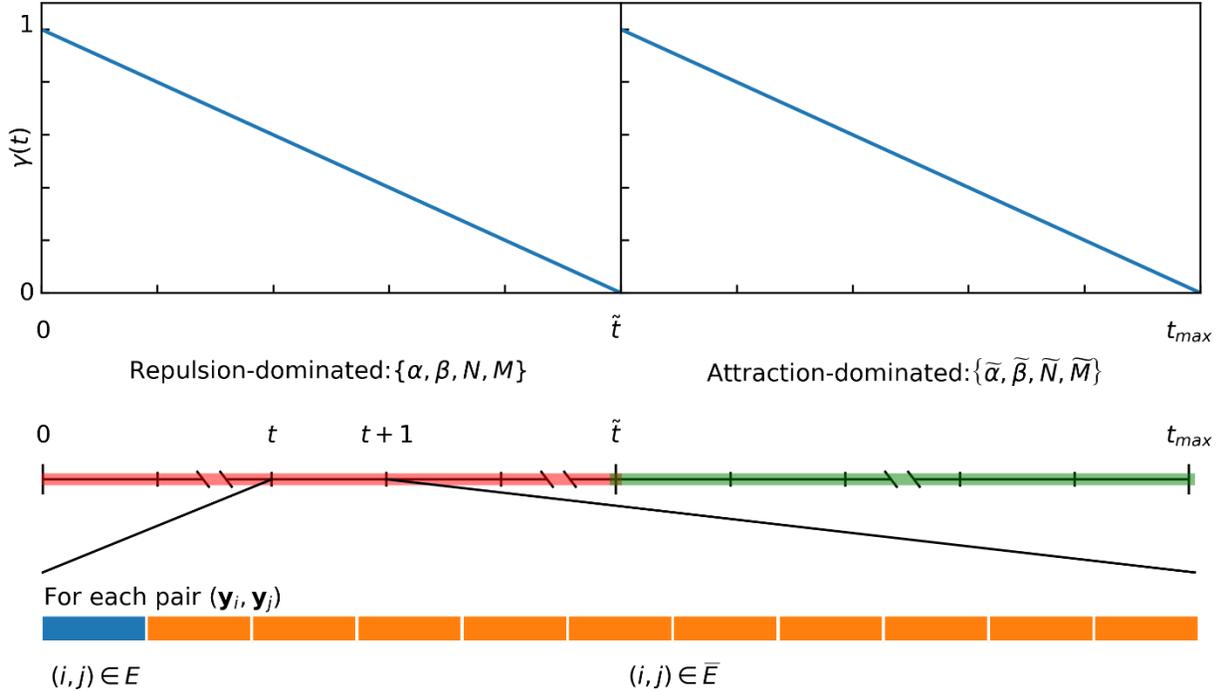

**Fig S11.** The schematic illustration of how forces are implemented during the two-stages of the iterative relaxed clustering. The red and green color bands show the repulsion-dominated and attraction-dominated stages respectively. Consider the iteration $Y(t) \to Y(t+1)$, for each feature

$y_i \in Y$ we apply the attractive forces between $y_i$ and its $k$-neighbors of the features (blue sub-bands), then apply repulsive forces between $y_i$ and randomly selected non-neighbors (orange sub-bands). The displacement of force relaxation is tuned by $\gamma(t)$, which starts from 1 in each stage, then linearly falls to nearly zero at the end of the stage.

4. Deriving the Gradient of UMAP Cost Function

UMAP introduces two fuzzy sets of input ($X$) and output data ($Y$), with two sets of weights being equivalent to $v_{ij}$ and $w_{ij}$ respectively. The divergence function for UMAP is denoted as the cross entropy of the two fuzzy set:

$$f_{UMAP} = \sum_{ij} \left[ v_{ij} log\left(\frac{v_{ij}}{w_{ij}}\right) + (1 - v_{ij}) log\left(\frac{1-v_{ij}}{1-w_{ij}}\right)\right].$$

Here, the input weights $v_{ij}$ are pre-calculated from $X$ and treated as constants. The output weights are given by,

$$w_{ij} = 1/(1 + ad_{ij}^{2b}),$$

where $d_{ij} = |y_i - y_j|$, $d_{ij} = d_{ji}$.

The gradient of the objective function with respect to $y$ is:

$$\frac{\partial f_{UMAP}}{\partial y_i} = 4 \sum_{j}^{N} \left[ abd_{ij}^{2(b-1)} v_{ij}/(1 + ad_{ij}^{2b}) - \frac{b(1-v_{ij})}{d_{ij}^2(1 + ad_{ij}^{2b})}\right] (y_i - y_j)$$

The two terms within the brackets in the last equation can be interpreted as attractive and repulsive forces acting on the features $y_i$ respectively.

Comparison of FR, t-SNE and UMAP

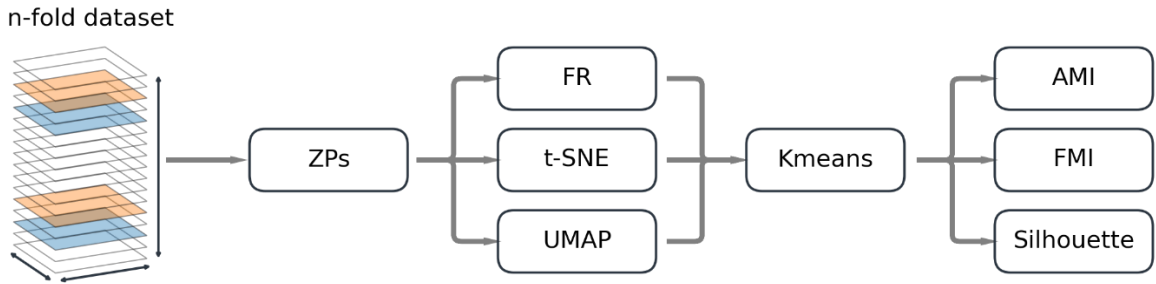

**Fig. S12. A workflow to compare t-SNE, UMAP and FR**. The synthetic patches consisting of two classes are represented by ZPs and embedded by t-SNE, UMAP and FR into a two-dimensional space. Kmean algorithm is applied to the corresponding 2D layout and AMI, FMI and Silhouette scores are evaluated.

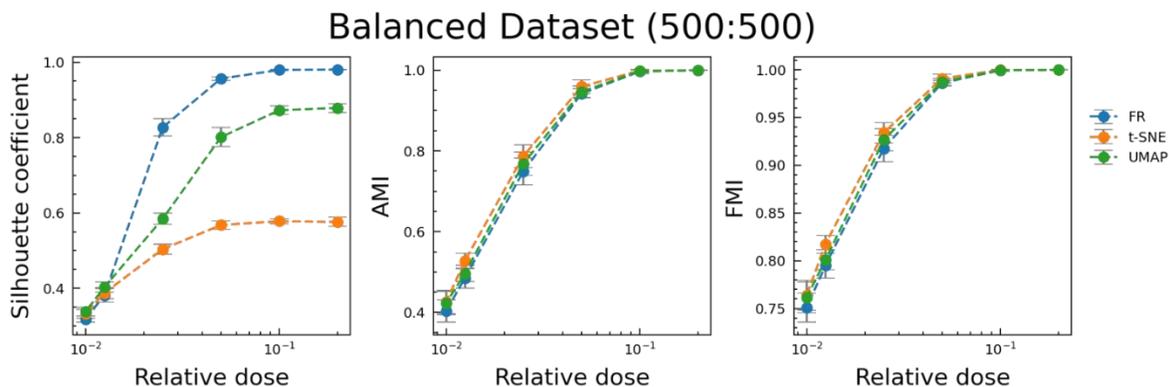

**Fig. S13. Comparison of FR, t-SNE and UMAP in synthetic balanced dataset.** From left to right: Silhouette coefficients, AMI, and FMI scores variation in the presence of Poisson noises.

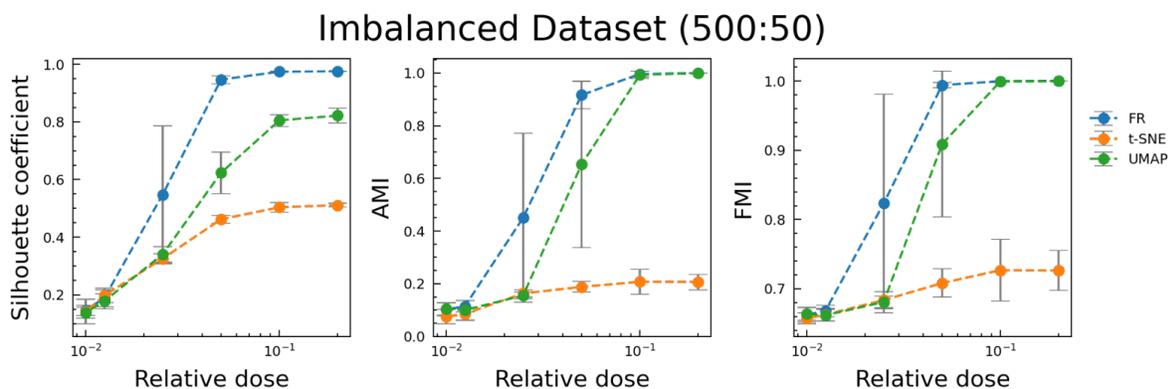

**Fig. S14. Comparison of FR, t-SNE and UMAP in synthetic imbalanced dataset.** From left to right: Silhouette coefficients, AMI, and FMI scores variation in the presence of Poisson noises.

We used clustering performance scores to evaluate and compare FR, t-SNE and UMAP. Different from a classification task, evaluating clustering performance is not as trivial as computing the precision and recall according to the labels. Specifically in the context of clustering, our metric should take the cluster separation into consideration rather than counting the absolute values of predicted labels. Similar to comparing different representation schemes (Fig. S8), we used

Silhouette coefficient, AMI and FMI to evaluate the performance of FR, t-SNE and UMAP (Fig. S13 and S14).

Following the workflow in Fig. S12, we found all three embedding techniques have comparable performance in the synthetic dataset with equal-sized clusters. In the uneven-sized cluster scenario, FR outperforms UMAP and t-SNE in Silhouette, AMI, and FMI scores.

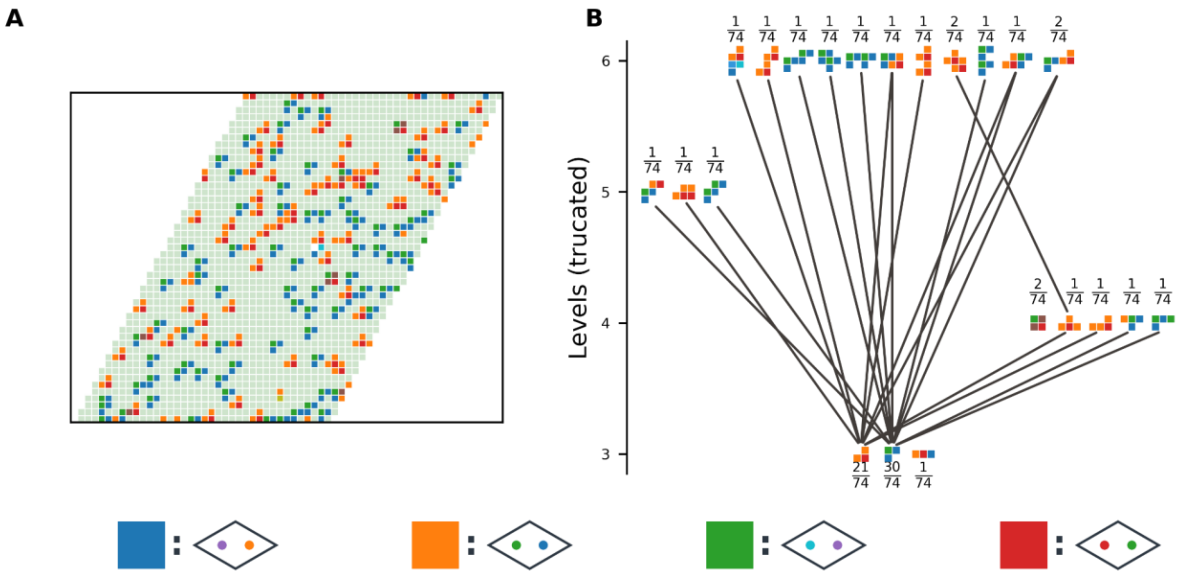

**Fig. S15. Construct a motif hierarchy of the WS$_2$ sample.** **(A)** Mapping realspace motif cells into a square grid. These motif-cells are colored according to the motif composition. (block legends below illustrate their realspace correspondents). **(B)** Motif-cells in (A) are ordered according to the number of cells. We associate higher-level motif-cells with lower-level ones if the spatial arrangement of cells in the latter occur within the former; here edges are drawn between associated motif-cells that are the nearest in the hierarchy.

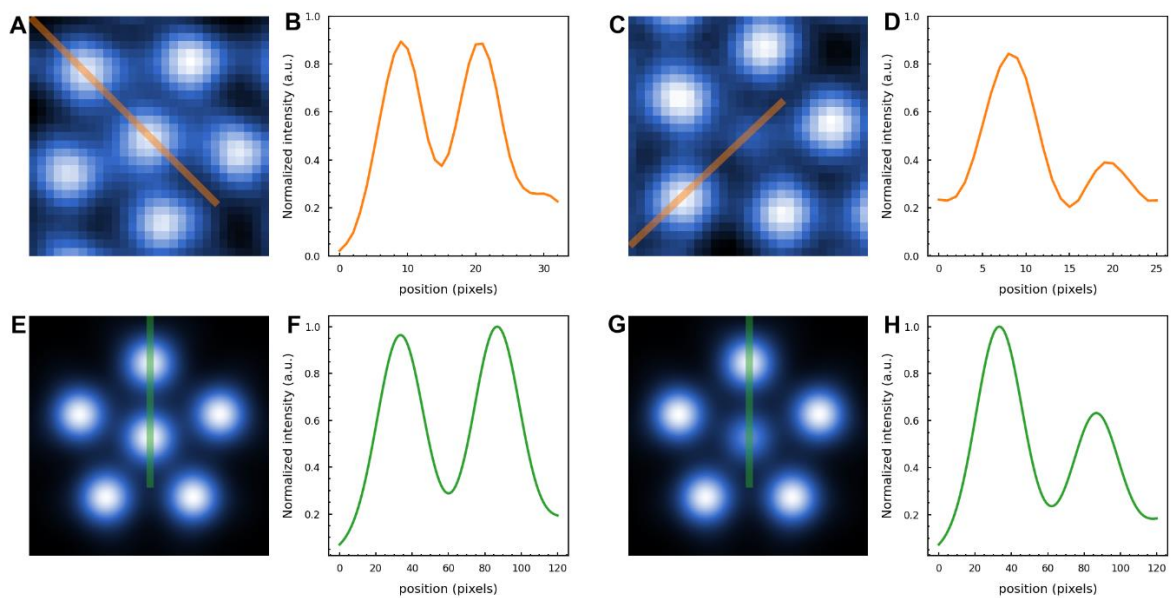

**Fig. S16. Line profile analysis of filled and partially filled pentagonal units in Figure 4 of main manuscript.** (A) Average motif of type1 filled pentagon unit. (B) Line profile extracted from (A), which is consistent with Mo or Ni centered pentagons. (C) Average motif of type 2 filled pentagon unit. (D) Line profile extracted from (C), which is consistent with V-centered motifs. (E) Simulation of filled pentagon unit. (F) Line profile extracted from (E). (G) Simulation of V filled pentagons unit. (H) Line profile extracted from (G).

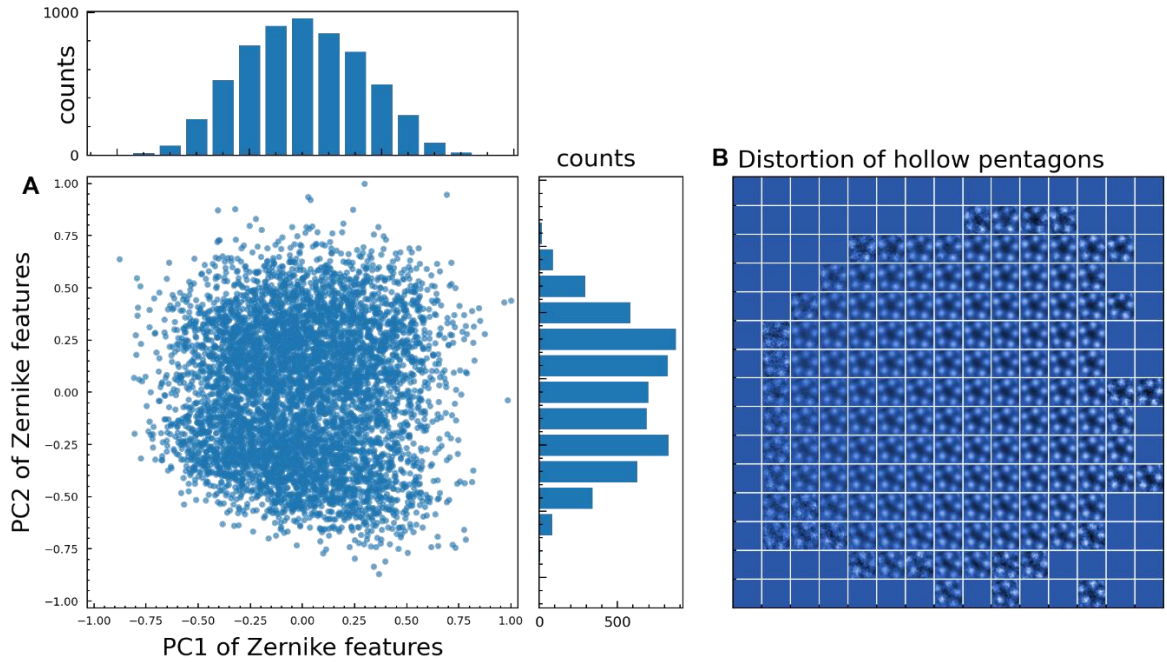

**Fig. S17. Distortion mapping of hollow pentagon units.** (A) PCA of rotational invariant Zernike features of hollow pentagon motifs. (B) Motif embedding showing the distortion of these hollow pentagon units.

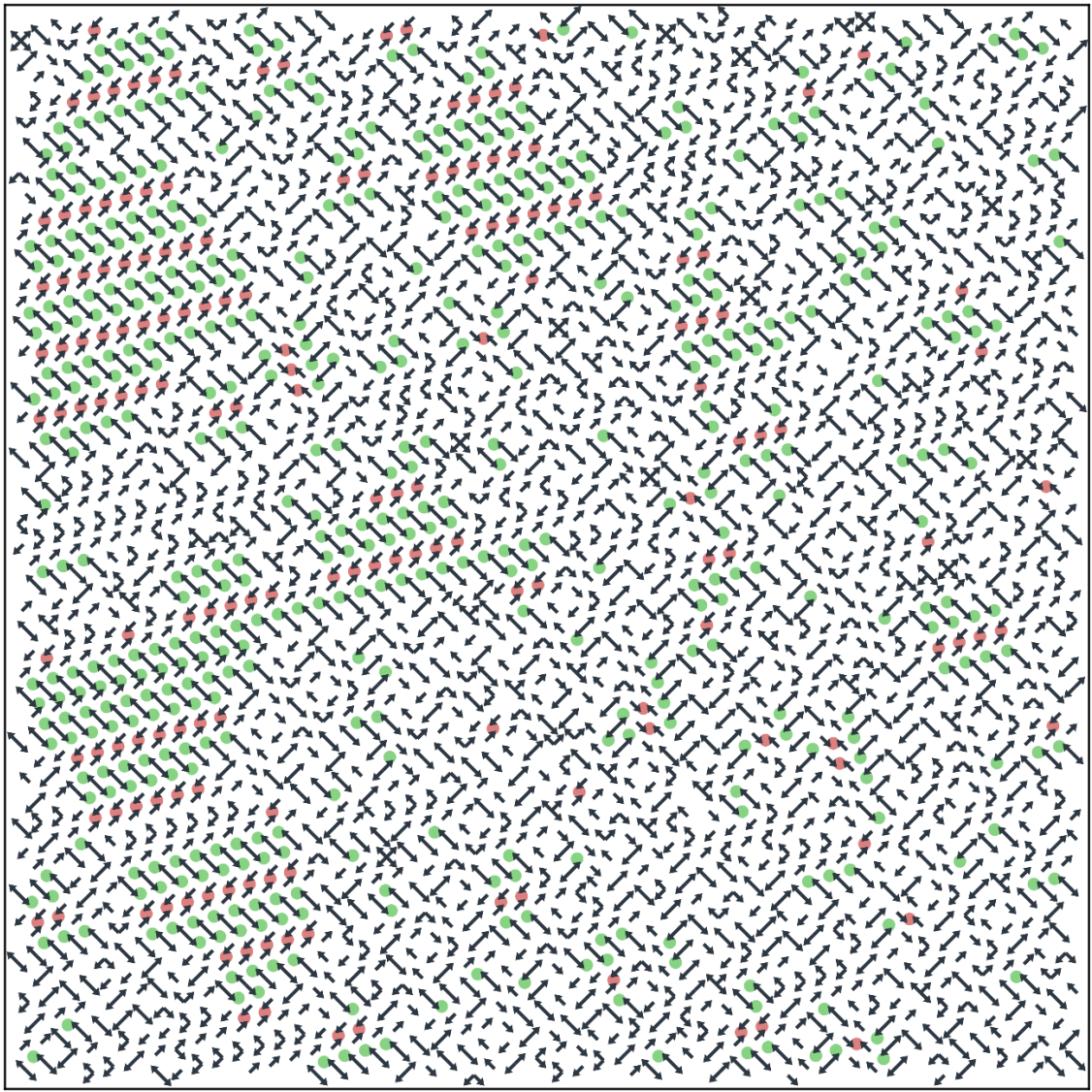

**Fig. S18. Reconstruction of the ADF-STEM image with arrows from second level motifs.**

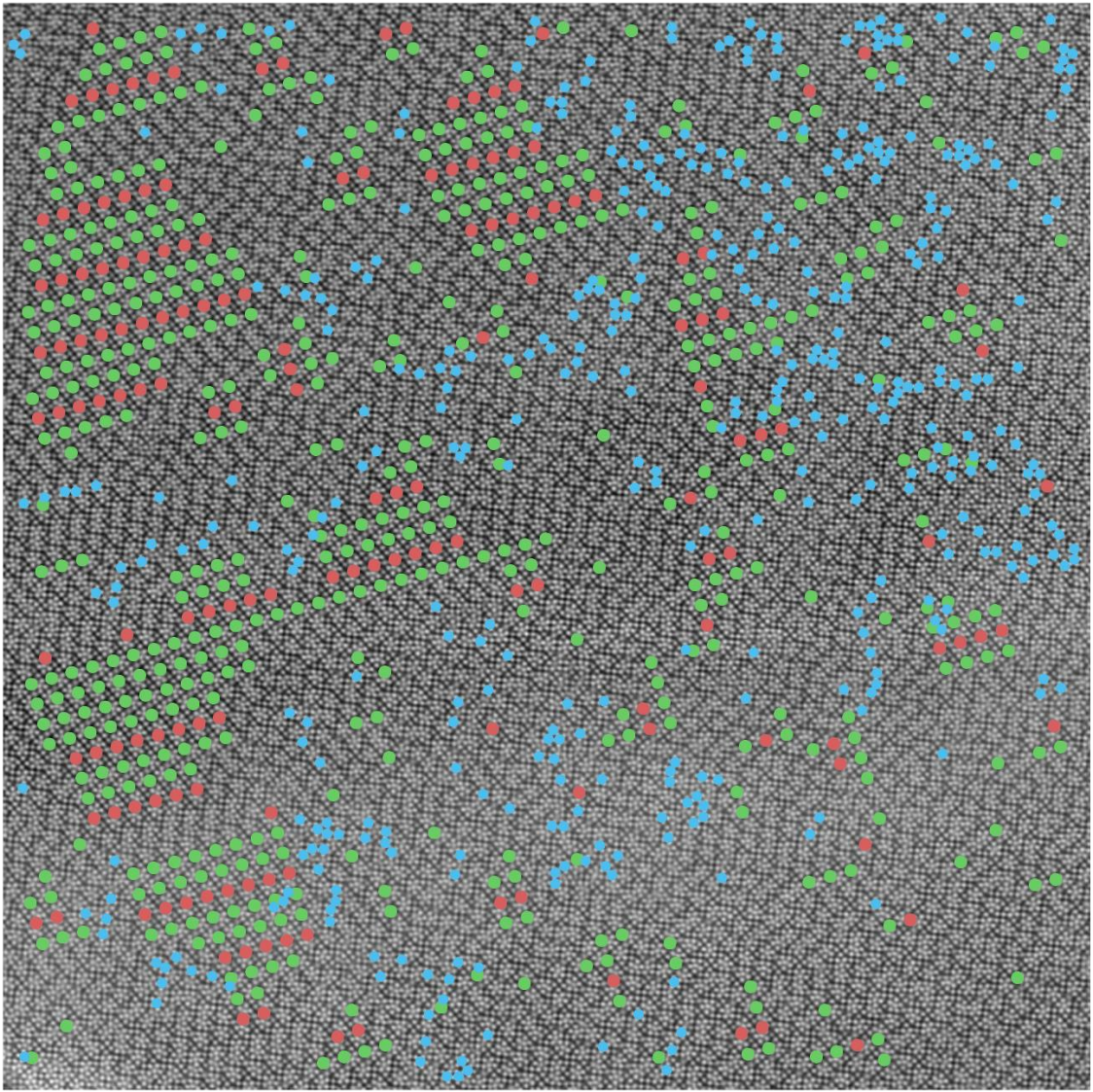

**Fig. S19. Largest dominant motifs clearly showing a novel phase that could tessellate the plane, but is frustrated by other competing structural motifs.**

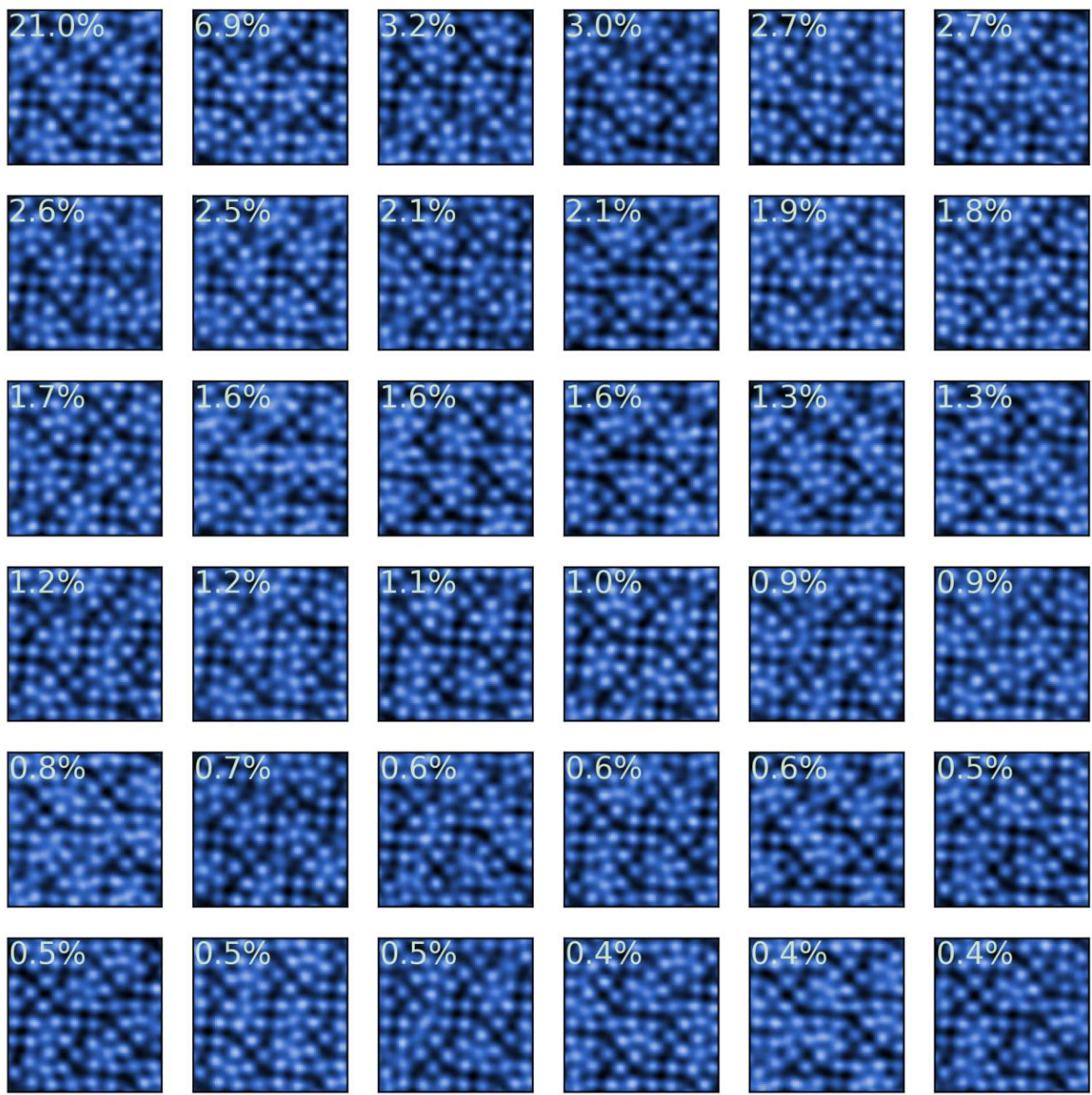

**Fig. S20.** Top 36 dominant third level motifs in the Mo-V-Te-Nb-oxide POM. They account for 74% of total third level motifs. The information entropy computed from all level 3 motifs via $-\sum p_i log(p_i)$ is 4.13, where $p_i$ is fraction of $i^{th}$ motif. The maximum information entropy in a completely disordered sample should be $log(2^{16}) \approx 11.09$.

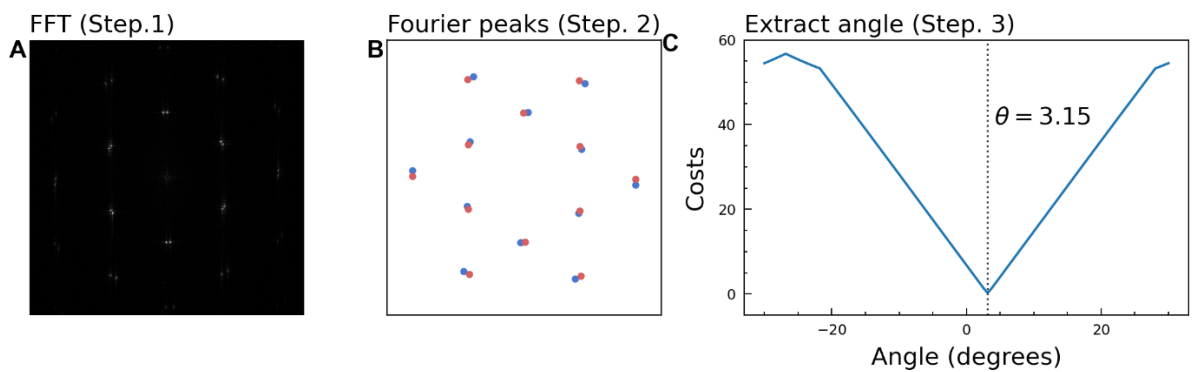

**Fig. S21. Workflow for extracting relative angle in bilayer MoS₂.** (A) the Fourier transformation of ADF-STEM image of bilayer MoS$_2$. (B) Extracted two set of peaks belonging to each layer of the moiré pattern. (C) By rotating one set of peaks in (B) and calculating the Euclidean distances mean to the other set of the peaks. The minimum value corresponds $\theta = 3.1$.

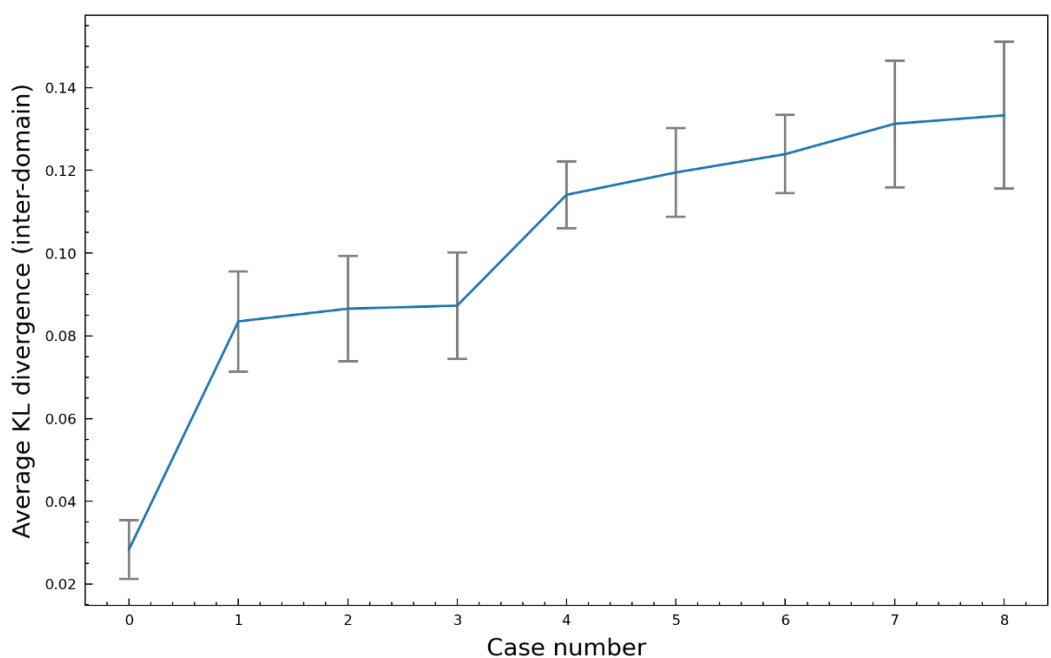

**Fig. S22. Evaluation of Kullback-Leibler (KL) divergences of difference domains in various cases of synthetic datasets.** Synthetic data is generated by adding two misaligned layers (with relative angle $\theta$) of hexagonal atomic columns which are modeled by two sets of Gaussian functions with relative intensity ratio $r$ and standard deviation $\sigma$ (also see Materials and Methods). Case 0: commensurate lattice $\theta = 3.14965734$. Case 1: incommensurate case $\theta = 3.2$. Case 2: increase the relative intensity ratio from 0.5 (case 1) to 0.8. Case 3: increase the size of Gaussian blobs from case 2 ($\sigma = \frac{a}{8} \rightarrow \sigma = \frac{a}{6}$). Case 4: add white noise ($\sigma_{noise} = 0.05$) to case 3. Cases 5 to 8: add random positional disorders to case 4 with maximum jump length setting to $0.02a$, $0.05a$, $0.1a$ and $0.15a$ respectively, where $a$ is the lattice constant of the hexagonal single layer of synthetic data.

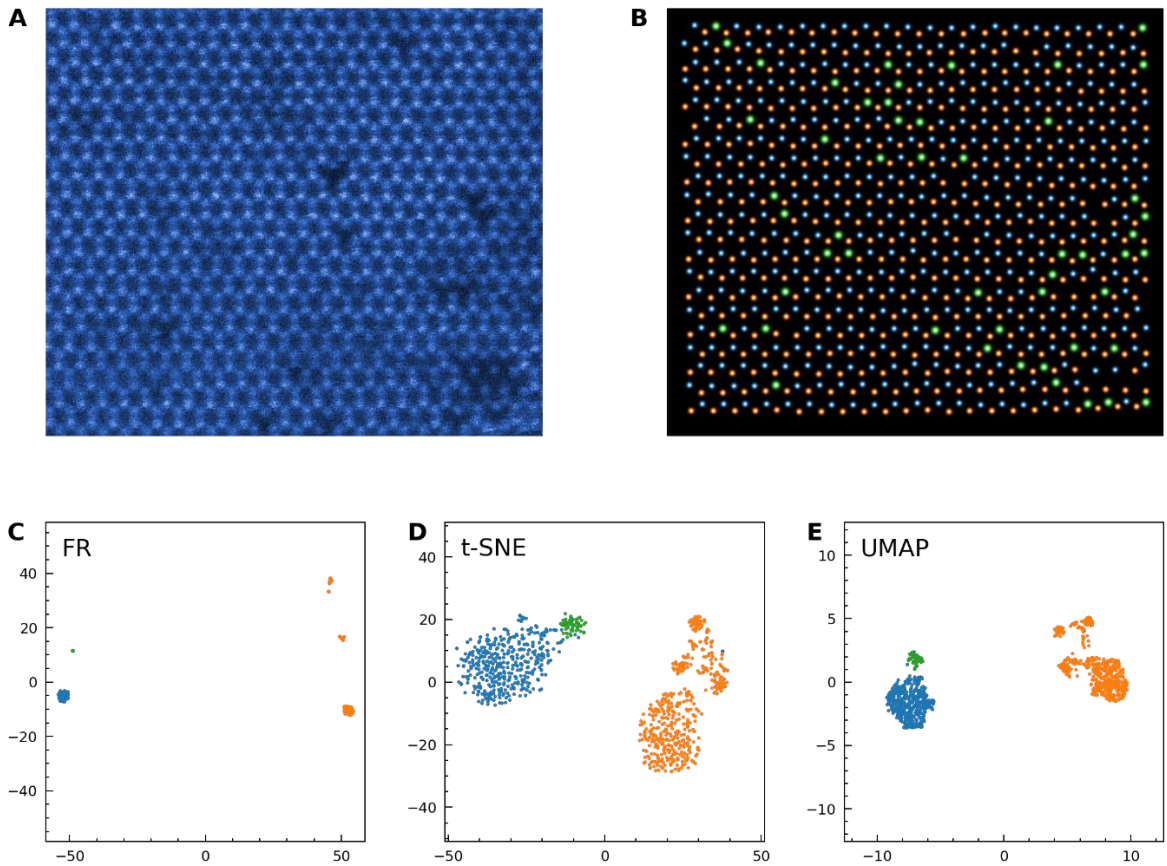

**Fig. S23 Identification of Se vacancies in a low SNR STEM image of monolayer MoSe$_2$ imaged with 40 kV electrons.** (a) ADF-STEM image of MoSe$_2$ with vacancy sites. (b) Identification map of Mo columns (orange), Se$_2$ columns (blue), and single Se vacancy columns (green) from (a). (c) Cluster map using the two-stage relaxed clustering scheme (this work). (d) Cluster map obtained from the t-SNE algorithm. (e) Cluster map obtained from the UMAP algorithm.

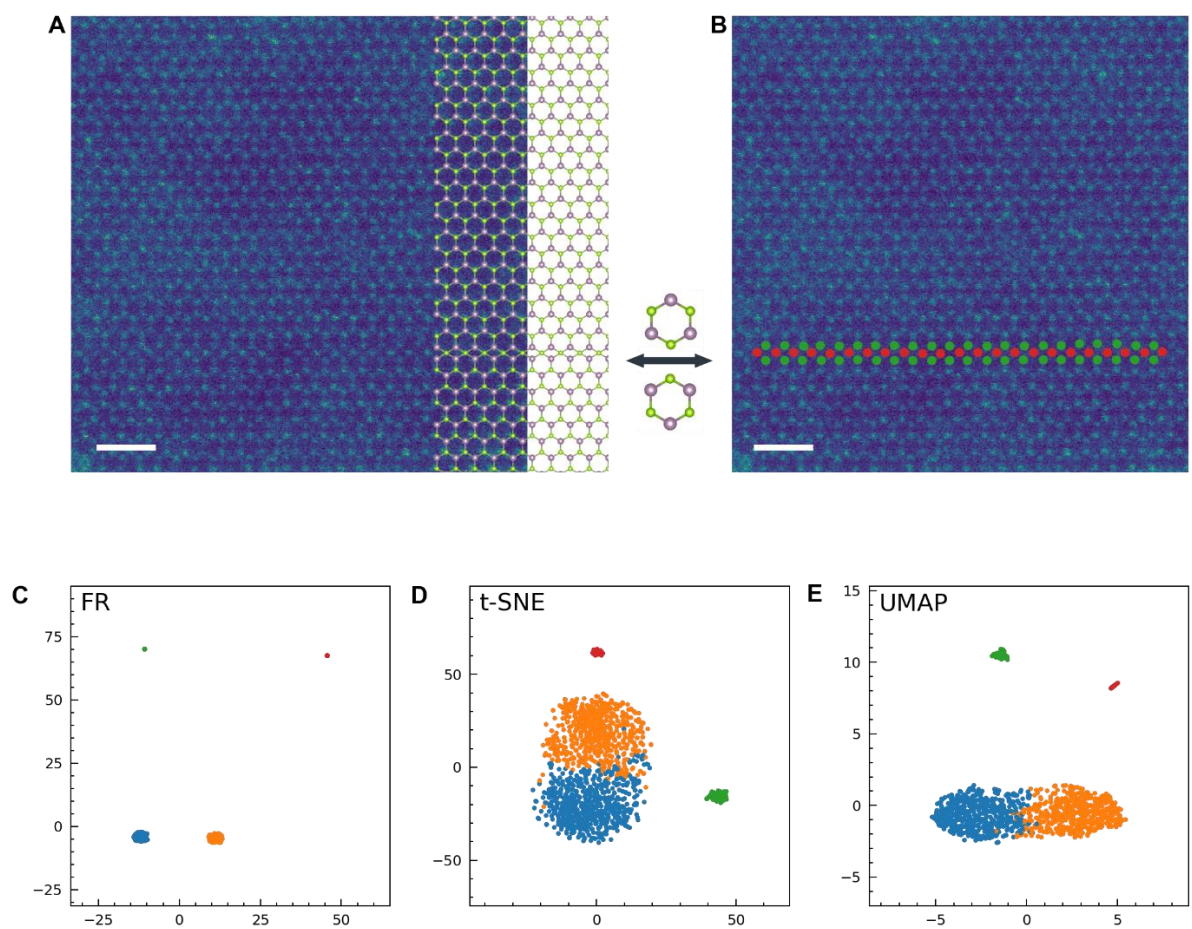

**Fig. S24 Identification of a mirror twin grain boundary (MTB) in low signal-to-noise ratio STEM image of monolayer MoSe$_2$. (A)** ADF-STEM image of monolayer MoSe$_2$ with MTB. **(B)** Identification of the grain boundary. The atomic columns indicated in green and red dots correspond to clusters of the same colour in **(C)**. **(C)** Cluster map obtained from the proposed two-stage relaxed clustering algorithm ($k$=10, $k'$=5). **(D)** Cluster map obtained from the UMAP algorithm. **(E)** Cluster map obtained from the t-SNE algorithm. The blue and orange clusters represent Mo columns and Se$_2$ columns respectively. (scale bar: 1nm)

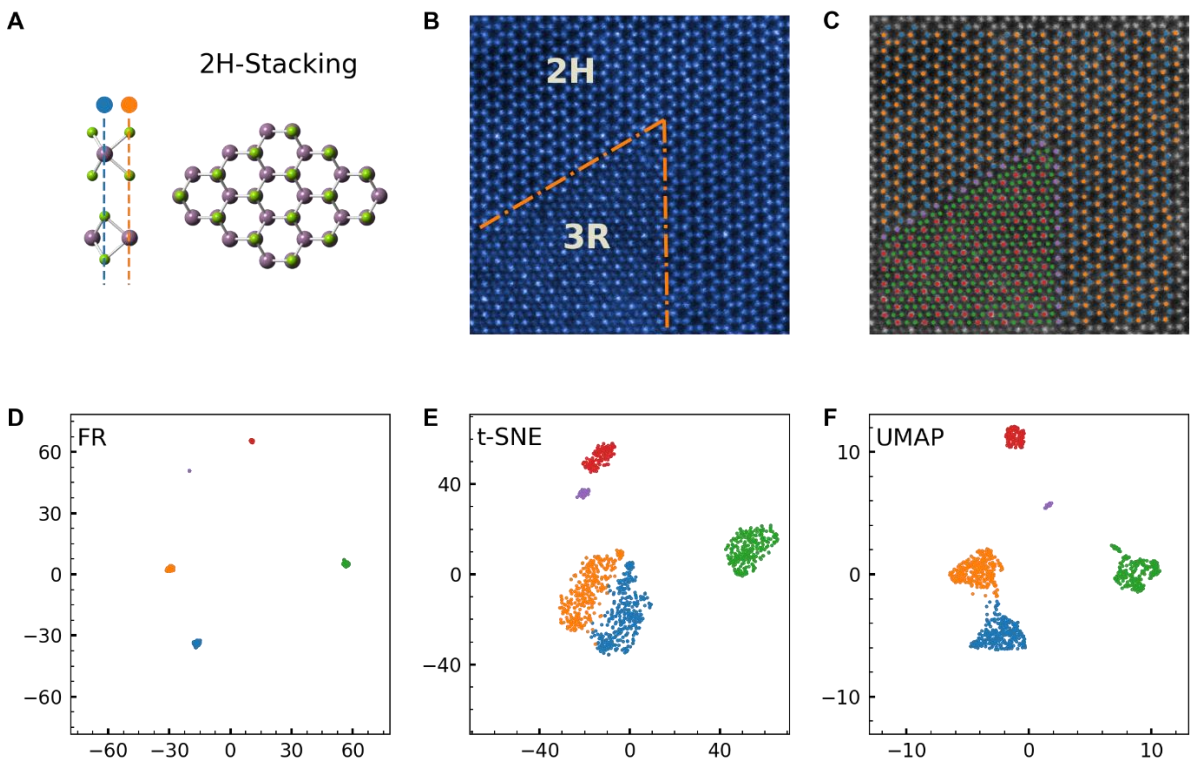

**Fig. S25 Identification of 2H and 3R phases and their phase boundaries in a STEM image of bilayer MoSe$_2$.** (a) Schematic model of 2H stacking. (b) ADF-STEM image of bilayer MoSe$_2$ with the 2H-3R boundaries indicated by orange dashed lines. (c) Identification map from the proposed method. (d) Cluster map obtained from the proposed two-stage relaxed clustering method ($k$=10, $k$'=5). (e) Cluster map obtained from t-SNE. (f) Cluster map obtained from UMAP.

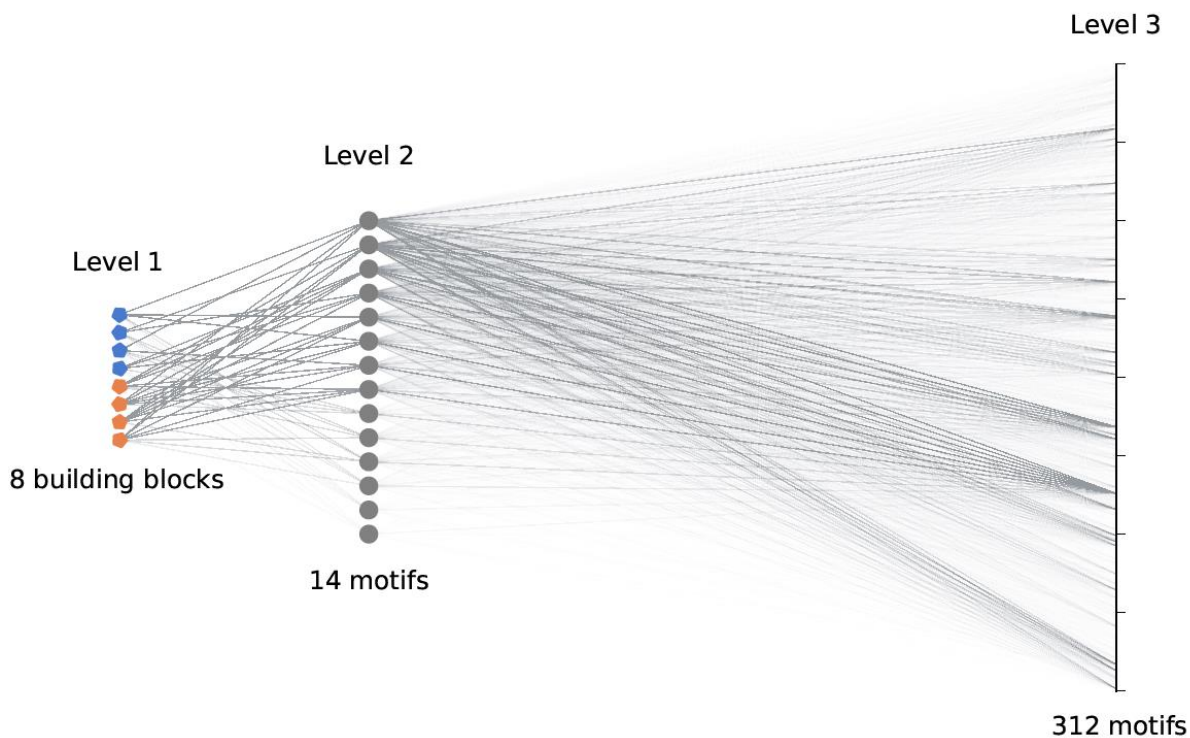

**Fig. S26 Complete motif hierarchy for Mo-V-Te-Nb-oxide POM**.

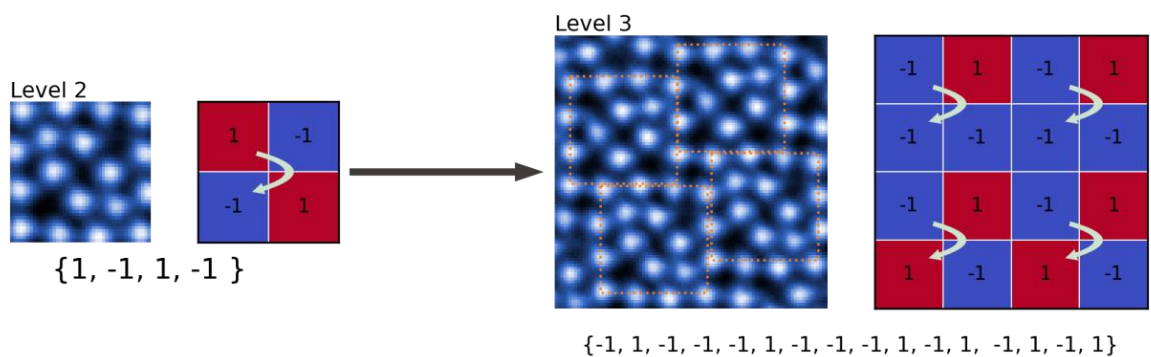

**Fig. S27. Encoding of level 2 and level 3 motifs using binary values of -1 and 1. -**1 denotes the hollow pentagon and 1 represents filled pentagon. Level 2 motifs can be encoded as a vector of length 4, and level 3 motifs can be represented by concatenating level 2 motif vectors, resulting in a vector of length 16.

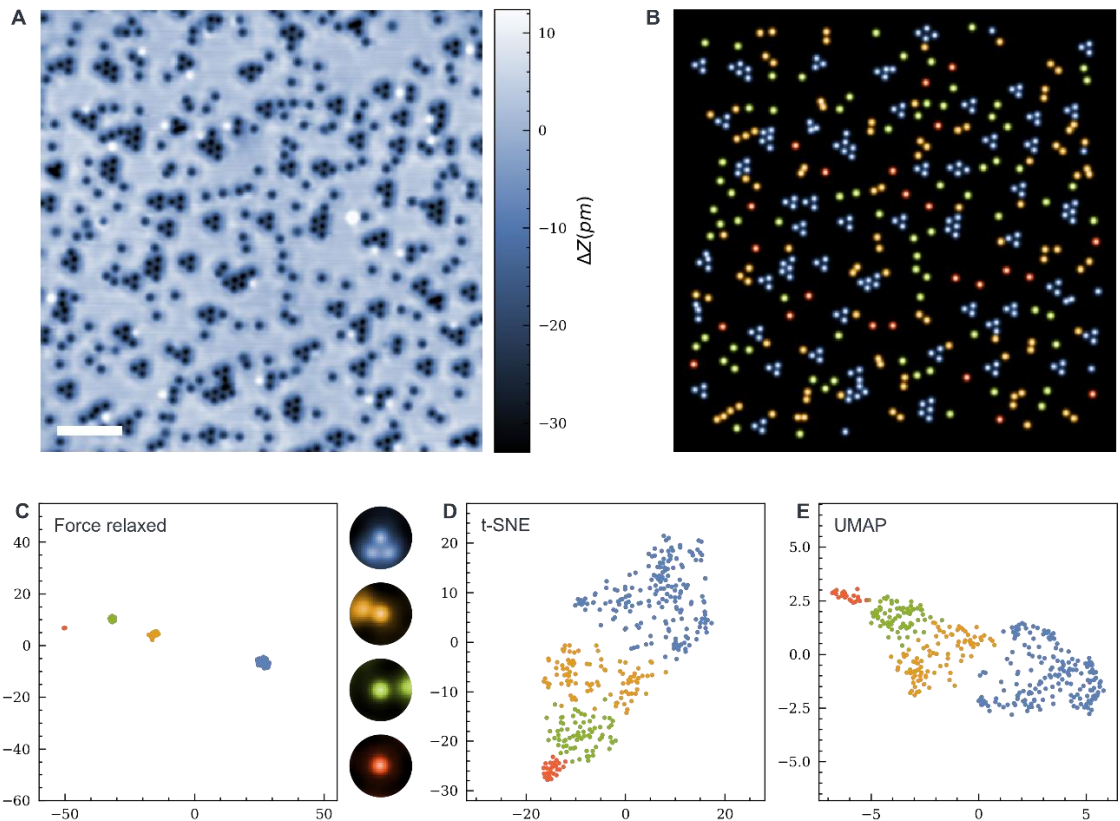

**Fig. S28. Comparison of clustering results using different low dimensional embedding methods in a STM image of a PtTe$_2$ thin film**. (**A**) STM image of PtTe$_2$ with Te vacancy sites (collected at sample bias $V_s$=-2 V and tunneling current $I$=300 pA) (**B**) Cluster map of all vacancy sites in (**A**). Red and blue dots correspond to isolated vacancies and clusters, while green and orange represent chains of different lengths. (**C**) Cluster map using a two-stage force-relaxation clustering scheme. (**D**) Cluster map obtained from the UMAP algorithm. (**E**) Cluster map obtained from the t-SNE algorithm. Scale bar: 5 nm in (A).

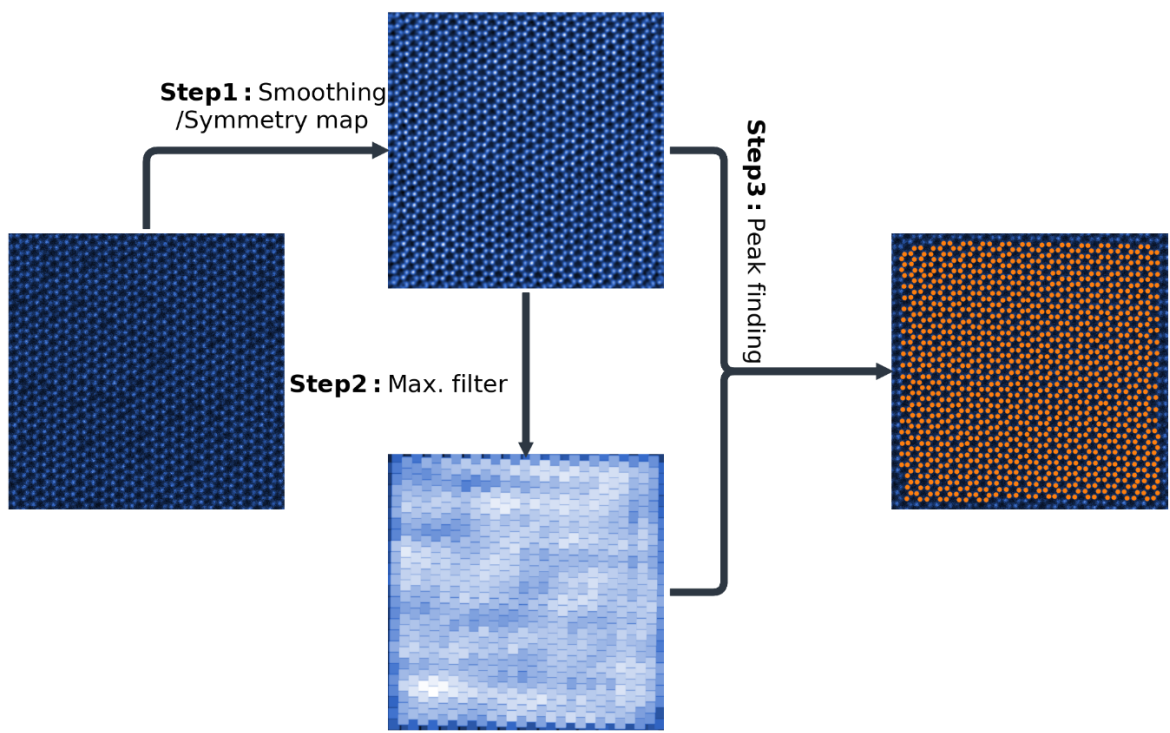

**Fig. S29. Workflow to locate the position of feature points**.

Fig. S29 illustrate a three-step workflow to locate feature points. The key to successfully extracting feature points is to obtain smooth versions of raw images (**Step 1**). Depending on image quality and image conditions, different smoothing schemes are adopted. For images with high signal-to-noise ratio (SNR), Fourier space filtering was implemented to keep 10% of lowest frequency components. For images with low SNR, Singular Value Decomposition based method was applied to the image. We adopted the SVD based method for images in this work unless stated otherwise. The smooth image was dilated by a local maximum filter (**Step 2**). The feature points are the locations where the input image is equivalent to the dilated version (**Step 3**).

Workflow to determine the patch size

Empirically the side length of patches is estimated to be the radius of the first dominant peak in radius distribution function (RDF) of the underlying sample, which can be approximated from Fourier transform of atomic resolution images. When the patch sizes are too large, the downstream FR algorithm will only capture the major clusters as small satellite clusters cannot be distinguished because they share high similarity with the corresponding major cluster; When the patch sizes are too small, the patch does not cover sufficient region and the local environment (spatial) information is lost, which make it comparable with conventionally Z-contrast method based on intensity profile.

Briefly, should the sample have a known periodicity, then the patch size should be about the same size as the repeating unit. This choice allows enough features to get discrete classes (i.e., motifs) from the FR clustering (see Figure S1). The residual disorder beyond the length scales of a single patch will then emerge in the hierarchy constructed from these motifs.

Here's our prescription for how patch sizes can be programmatically determined. When it comes to the selection of patch sizes, it is a hyperparameter that can be automatically computed via the following steps:

1. Compute the FFT of the input image (power spectrum image).

2. Obtain the average radial intensity curve from the power spectrum image and locate the radius of the first dominant peak from the average radial intensity curve.

3. Convert the radius calculated from FFT space to image (real) space.

The above process is summarized in a workflow as shown in Fig. S30.

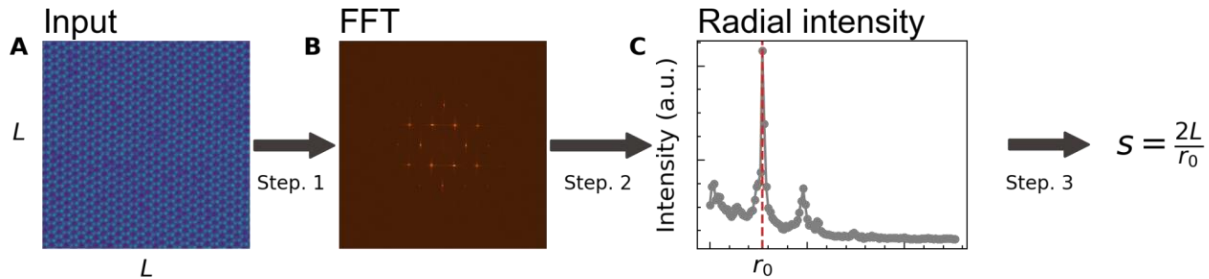

**Fig. S30. Workflow to estimate the motif size for patch extractions.** (**A**) An atomic resolution STEM image with a shape of (L, L). (B) FFT of input image in (A). (**C**) Average radius intensity calculated from panel (B), and $r_0$ is the position of the first dominant peak. In the final step, the patch size is empirically calculated by $s = \frac{2L}{r_0}$.

| (p, q) | (0, 0) | (1, -1) | (1, 1) | (2, -2) | (2, 0) | (2, 2) | (3, -3) | (3, -1) | (3, 1) |
|---|---|---|---|---|---|---|---|---|---|
| j | 0 | 1 | 2 | 3 | 4 | 5 | 6 | 7 | 8 |
| (p, q) | (3, 3) | (4, -4) | (4, -2) | (4, 0) | (4, 2) | (4, 4) | (5, -5) | (5, -3) | (5, -1) |
| j | 9 | 10 | 11 | 12 | 13 | 14 | 15 | 16 | 17 |

**Table S1. Relation between indices $(p, q)$ and $j$**

| **FR Clustering** | |
|---|---|
| **Total iterations** | Time (s) |
| **10** | 0.57 ± 0.05 |
| **20** | 0.70 ± 0.05 |
| **40** | 0.87 ± 0.05 |
| **80** | 1.28 ± 0.10 |
| **160** | 2.17 ± 0.12 |
| **Hierarchy Construction** | |
| Step 1. Identify motif-cells | 0.17 ± 0.03 |
| Step 2. Connect motif-cells | 0.10 ± 0.02 |
| Step 3. Construct hierarchy | 0.32 ± 0.01 |

**Table S2**. Computation time for FR clustering for various number of iterations ($k$=15, $k$'=5, PCA initialization). These timing tests were performed on 1207 features from the model system monolayer MoSe$_2$ in Fig. 1A, using a single-threaded process on a desktop computer with Intel Xeon CPU E5-2630Lv3 @ 1.80 GHz and 24 GB RAM. Given the same hardware, concurrent programming paradigms will further accelerate these processing times.

| | Class 1 ($\sigma = 7$) | Class 2 ($\sigma = 7$) |
|---|---|---|
| 3-fold | 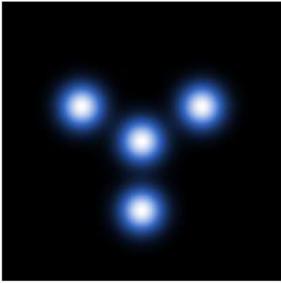 | 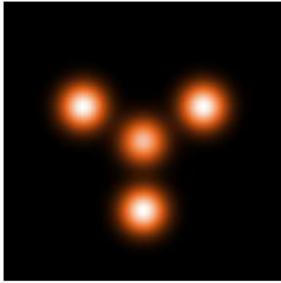 |
| 4-fold | 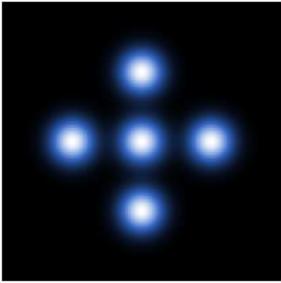 | 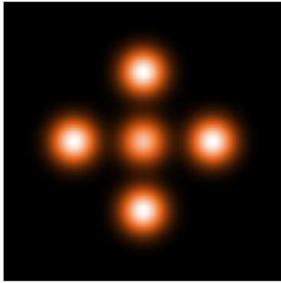 |
| 5-fold | 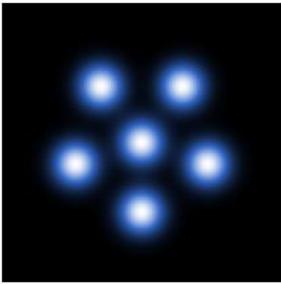 | 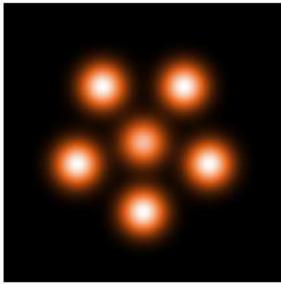 |
| 6-fold | 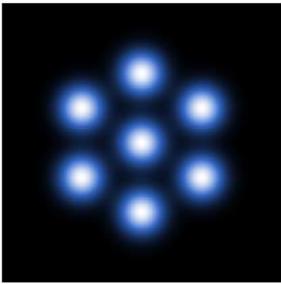 | 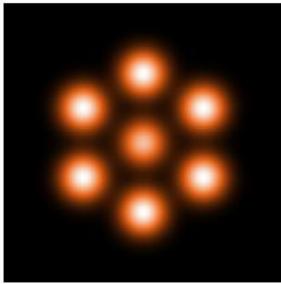 |
| 7-fold | 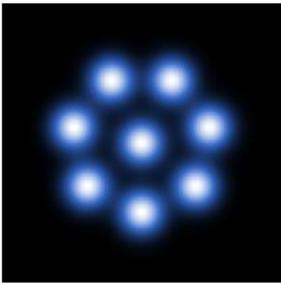 | 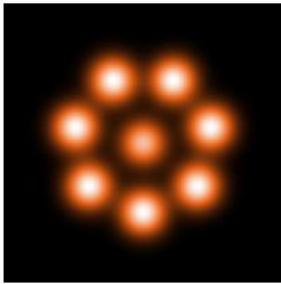 |

**Table S3**. A visual display of the synthetic dataset when default parameters are used.

**Synthetic Dataset description**: By default, each synthetic motif with *n*-fold rotational symmetry has a shape of 128x128 which contains (*n*+1) Gaussian blobs. These (*n*+1) Gaussian blobs form a regular n-gon with the extra Gaussian blob occupying the center. The surrounding Gaussian blobs have a distance of 32 pixels away from the center of the motif. All Gaussians share the same sigma value of 7. In all blue motif classes, all Gaussian blobs have a maximum amplitude of 1.0, while in the orange motif classes, the central blob has a maximum amplitude of 0.80. In case of applying Poisson noise to the above data, the motif is scaled and converted to integer type.

A synthetic dataset can be specified by a unique set of parameters $\{n1, n2, s, \sigma, l, A\}$, and the definitions are these parameters are declared as follows (default values in parentheses):
- $n1$ – the number of patches for class 1
- $n2$ – the number of patches for class 2
- $s$ – the size of every patch ($s = 128$)
- $\sigma$ – the sigma value of the Gaussian blobs contained in each patch ($\sigma = 7$)
- $l$ – the distance of surrounding Gaussian blobs to the central blobs ($l = 32$)
- $A$ – the relative intensity of central blobs in class 2 ($A = 0.8$)

| | **Attraction force** | **Repulsion force** | **Remarks** |
|---|---|---|---|
| LargeVis | $\dfrac{2}{(1+d^2)}$ | $\dfrac{2\gamma}{(1+d^2)(d^2+\epsilon)}$ | $\gamma = 7, \epsilon = 0.1$ |
| UMAP | $\dfrac{2ab\, d^{2b-2}}{(1+a\, d^{2b})}$ | $\dfrac{2\gamma b}{(1+a\, d^{2b})(d^2+\epsilon)}$ | $\gamma = 1, \epsilon = 0.001$ $a = 1.58, b = 0.90$ |
| **Two-stage relaxed clustering (this work)** | $\dfrac{\alpha}{1+d^n}$ | $\dfrac{\beta}{1+d^m}$ | $m \geq 0, n \geq 0$ $\alpha > 0, \beta > 0$ |
| Here $d$ refers to the distance between two PCA-reduced features: $\boldsymbol{d} = ||\boldsymbol{Y_u} - \boldsymbol{Y_v}||$. | | | |

**Table S4. A summary of force functions in different iterative clustering algorithms**